\documentclass[letterpaper,twocolumn,10pt]{article}
\pdfoutput=1
\PassOptionsToPackage{hyphens}{url}\usepackage{hyperref}

\usepackage{style/usenix2019_v3}
\usepackage[available,functional,reproduced]{style/usenixbadges}
\microtypecontext{spacing=nonfrench}

\usepackage{soul}
\usepackage{listings}
\usepackage{placeins}

\usepackage{graphicx}
\usepackage{tabularx}
\usepackage{booktabs}
\usepackage[nointegrals]{wasysym}
\usepackage{etoolbox}

\usepackage{enumitem,amssymb}
\newlist{todolist}{itemize}{2}
\setlist[todolist]{label=$\square$}
\usepackage{pifont}

\usepackage[makeroom]{cancel}

\usepackage{threeparttable}
\usepackage{makecell}

\usepackage{graphicx}
\usepackage{subcaption} 
\usepackage{comment}
\usepackage{enumerate}
\usepackage{xspace}

\usepackage{multirow}
\usepackage{soul}

\usepackage{amsthm}
\usepackage{amsfonts}
\usepackage{mathtools}
\usepackage[utf8]{inputenc}

\usepackage{color}
\usepackage{xcolor}

\usepackage{enumitem}
\setlist[itemize]{noitemsep, topsep=0pt, leftmargin=10pt}
\setlist[enumerate]{noitemsep, topsep=0pt, leftmargin=10pt}

\usepackage{colortbl}
\definecolor{Gray}{gray}{0.65}
\definecolor{LightGray}{gray}{0.9}

\usepackage{array}
\usepackage{arydshln}
\newcommand{\lsec}[1]{\label{sec:#1}}
\newcommand{\lfig}[1]{\label{fig:#1}}
\newcommand{\ltab}[1]{\label{tab:#1}}

\newcommand{\rsec}[1]{\S\ref{sec:#1}}
\newcommand{\rfig}[1]{Figure~\ref{fig:#1}}
\newcommand{\rtab}[1]{Table~\ref{tab:#1}}

\newcommand{\comp}[1]{\textit{#1}}

\newcommand{\secspacingtop}{\vspace{-0pt}}
\newcommand{\secspacingbot}{\vspace{-0pt}}
\newcommand{\subsecspacingtop}{\vspace{-0pt}}
\newcommand{\subsecspacingbot}{\vspace{-0pt}}

\newcommand{\sectionpage}{}

\usepackage{titlesec}

\newcommand{\oursystem}{Zeph\xspace} %

\newcommand{\fakeparagraph}[1]{\vskip 0pt\noindent\textbf{#1 }}

\makeatletter
\newcommand{\printfnsymbol}[1]{%
  \textsuperscript{\@fnsymbol{#1}}%
}
\makeatother

\newcommand{\funcs}{$\Sigma_S$\xspace} %
\newcommand{\funcm}{$\Sigma_M$\xspace} %
\newcommand{\funcdp}{$\Sigma_{DP}$\xspace} %

\begin{document}

\title{
\Large \bf \oursystem: Cryptographic Enforcement of End-to-End Data Privacy\hypersetup{linkcolor=black}\thanks{This is an extended version of a paper published at OSDI 2021.}}

\patchcmd{\maketitle}
	{\@maketitle}
	{\vspace{-0em}\@maketitle\vspace{-0em}}%
	{}
	{}

\author{
{\rm Lukas Burkhalter\hypersetup{linkcolor=black}\thanks{These authors contributed equally to this work.}, Nicolas Küchler\footnotemark[2], Alexander Viand, Hossein Shafagh, 
Anwar Hithnawi}  \\ 
\\ 
{\textit{ETH Z{\"u}rich}}  %
}

\date{}

\maketitle
\pagestyle{empty}

\begin{abstract}

As increasingly more sensitive data is being collected to gain valuable insights, the need 
to natively integrate privacy controls in data analytics frameworks is growing in importance. 
Today, privacy controls are enforced by data curators with full access to data in the clear.
However, a plethora of recent data breaches show that even widely trusted service providers can be compromised.
Additionally, there is no assurance that data processing and handling comply with the claimed privacy policies.
This motivates the need for a new approach to data privacy that can provide strong assurance and control to users.
This paper presents \oursystem, a system that enables users to set privacy preferences on how their data can be \textit{shared} and \textit{processed}.
\oursystem enforces privacy policies cryptographically and ensures that data available to third-party applications complies with users' privacy policies.
\oursystem executes privacy-adhering data transformations in real-time and scales to thousands of data sources, 
allowing it to support large-scale low-latency data stream analytics.
We introduce a hybrid cryptographic protocol for privacy-adhering transformations of encrypted data.
We develop a prototype of \oursystem on Apache Kafka to demonstrate that \oursystem can 
perform large-scale privacy transformations with low overhead.

\end{abstract}

\subsecspacingtop
\section{Introduction}
\subsecspacingbot
\lsec{introductions}

The availability of rich data and the advancement of tools and algorithms to process data at scale has enabled tremendous innovations in various fields ranging from health and retail to agriculture and industrial automation~\cite{philipscovid19,oracleretail,innovationagriculture}.
However, the accumulation of sensitive data has made service providers hosting data lakes a desirable target for attacks. 
In addition, a surge of incidents of unauthorized data monetization, instrumentation, and sharing has raised societal concerns~\cite{amnesty, surveillance}.
This has pushed regulatory bodies to enact data privacy regulations
 to prevent misuse of private data and ensure the privacy of personal data~\cite{gdpr,ccpa}.
Today, the most integral parts of existing data protection systems are security controls 
such as authentication, authorization, and encryption
which protect data by guarding it and limiting unnecessary exposure.
Security controls alone, however, are not sufficient. 
 We ultimately need to ensure that user's privacy is respected even by entities authorized to use the data.
Thus, privacy solutions that control the extent of what can be \emph{inferred}~\cite{dp_uscensus} from data and protect \emph{individuals' privacy}~\cite{erlingsson2014rappor} are crucial if we are to continue to extract utility from data safely.

\fakeparagraph{Today's Data Privacy Landscape:}
The advent of new data privacy regulations such as GDPR and CCPA, coupled with the increasing importance of data,
has led to a growing demand for privacy solutions that protect sensitive data while retaining its value. 
Despite recent advancements in privacy enhancing technologies~\cite{Dwork2006-mp,wrk, generalizationkanonimity}, %
privacy frameworks~\cite{riverbed, privay-policies-if, thoth-if, sgx-if, ifc-sgx-pets, taint-droid-adnoird-ifc} remain shaped by regulatory requirements that predominately focus on the notion of \emph{notice and consent}~\cite{nytimes,immuta, privitar}.
 Though an essential step towards transparency and user control, it is important to emphasize that user consent is not the answer to data privacy.
  Bad practices in data use and sharing remain pervasive in consent-based systems~\cite{consent1,consent2,consent3}, and often consent does not adequately express the complexities of real-world privacy preferences.
The status quo has three shortcomings that we aim to address with this work:
\emph{(i)~Trusted data curators:} In the current model, privacy controls are implemented and enforced by data curators 
who have full access to data in the clear.
Frequent data breaches~\cite{techcrunch, capital-one, enterprise-data-breach} have shown that even trusted providers can be compromised or fall prone to data misuse temptations.
Additionally, there are no assurances that data processing actually complies with the stated privacy policies.
Consequently, there is a need for built-in data privacy mechanisms that do not require data curators to access data in the clear.
\emph{(ii)~Lack of user control:}
Though privacy regulations mandate services to grant users more control over their data,
the materialization of this has been disappointing in practice. %
Services have been drafting privacy policies that unilaterally dictate how users' data will be used.
Users have no option to exert their \mbox{data privacy} preferences except to give blanket consent if they choose to use the service~\cite{nytimes,theguardian}.
\emph{(iii)~End-to-end privacy:} 
Privacy solutions today are mostly ad hoc efforts~\cite{Gartner} rather than an integral part of the data processing ecosystem.
We need a cohesive end-to-end approach to data privacy that follows data from source to downstream.
Such solutions should integrate with existing data processing and analytics frameworks and coexist with data protection mechanisms \mbox{already in place}.

\fakeparagraph{\oursystem:}
In this work, we propose \oursystem, a new data privacy platform that provides the means to safely extract value from encrypted data 
while ensuring data confidentiality and privacy by serving only privacy-compliant data. 
\oursystem addresses the above shortcomings with two key ideas:
\textit{(i)}~a user-centric privacy model that enables users to express their privacy preferences.
In \oursystem, a user can authorize services to access raw data or privacy-compliant data securely.
This aligns with data sharing practices claimed in privacy policies today: e.g., "we share or disclose your personal data with your consent" or "we only provide aggregated statistics and insights"~\cite{twitter-privacypolicy, instagram-privacypolicy}.
In addition to this commonly referenced aggregation policy, 
\oursystem supports more advanced privacy-compliant data transformations.
For example, transformations that restrict what can be inferred from the data (e.g, generalization techniques~\cite{generalizationkanonimity, generalization_websearches, generalization_medical}) or ensure differential privacy --~a mathematically rigorous definition of privacy.
\textit{(ii)}
\oursystem cryptographically enforces privacy compliance and executes privacy transformations on-the-fly over encrypted data, ensuring that the generated transformed views \mbox{conform to users' privacy policies}. 

The design concepts underpinning \oursystem are generic and could be adapted to other systems. 
In this work, we specifically target data stream analytics/processing pipelines and build on the typical structure of such systems. 
Hence, we focus on cryptographic building blocks that optimize efficiency for this type of data.
Streaming compute tasks are increasingly relevant in various privacy-sensitive sectors~\cite{google-stream-processing,appache-flink, twitter-heron, noria}.
The online nature of stream processing makes low latency and high throughput critical requirements for privacy-preserving stream processing solutions.

\fakeparagraph{Cryptographically Enforced Privacy Transformations.} 
There are three key challenges in designing a data platform that enables privacy-compliant data transformations on encrypted data. 
First, we need to ensure compatibility with the data flow of existing data processing pipelines (e.g., storage and compute) and meet their strict performance requirements.
Second, the platform must enable a wide range of existing privacy transformations and allow for different transformations to be applied to the same underlying data.
Finally, in addition to single-source privacy transformations, we need to support transformations that require combining data from multiple users (e.g., aggregate private data releases).

 Existing practical encrypted data processing systems generally use partially homomorphic encryption schemes that already support the single-source privacy transformations required in our system~\cite{timecrypt,seabed,cryptdb,monomi,prio,ml_encoding}.
However, homomorphic evaluation alone is insufficient to support aggregations across data from different users.
 Supporting these functions is typically achieved via multi-party computation protocols that are optimized for aggregation operations~\cite{prio,sketch,castelluccia2011CancelOut,honeycrisp-secure-aggr}.
 These protocols ensure that user inputs remain private and only the aggregation result is revealed to the server.
 However, these protocols are either limited to specific functions (e.g., updating sketches) or require the data producers to take an active part in the computation.

We address these challenges in \oursystem using two ideas:
\textit{(i)}~a new approach for encryption that decouples data encryption from privacy transformations. 
This logical separation of the data and privacy plane allows us to remain compatible with data flows in existing systems.
Data producers remain oblivious to the transformations and do not need to encrypt data towards a fixed privacy policy.
\textit{(ii)}~ we introduce the concept of cryptographic \emph{privacy transformation tokens} to realize flexible data transformations.
These tokens are, in essence, the necessary cryptographic keying material that enables the respective transformation on encrypted data.
\oursystem creates these tokens via a hybrid construction of secure multi-party computation (MPC) and a partially homomorphic 
encryption scheme. 
Outputs of privacy transformations over encrypted data at the server-side are then released by combining the encrypted data with corresponding cryptographic \mbox{transformation tokens}.

We have built a prototype of \oursystem\footnote{\oursystem's code available at: \url{https://github.com/pps-lab/zeph-artifact}}that is interfaced with Apache \emph{Kafka}~\cite{kafka-online}. 
Our evaluation results show that \oursystem can serve real-time privately transformed streams in different applications
with a 2x to 5x latency overhead compared to plaintext.
We optimize the interactive part of the underlying MPC protocol with ideas from graph theory to achieve the scalability requirements of \oursystem. Our optimization improves performance up to 55x compared to the baseline.

\sectionpage
\secspacingtop
\section{Overview}
\secspacingbot
\lsec{overview}
\subsecspacingtop
In this section, we discuss end-to-end privacy and its requirements, give an overview of \oursystem, and describe our security and privacy model.

\subsecspacingtop
\subsection{End-to-End Privacy}
\subsecspacingbot
\lsec{privacy-transformations}
In this work, we investigate a new cohesive end-to-end design for data privacy.
Despite being heavily intertwined with users' data, data systems have evolved with design objectives centered around availability, performance, and scalability, while privacy is essentially overlooked.
As privacy becomes a more urgent concern, we need system designs that retrofit privacy into existing established data framework designs. 
Embedding privacy in the current complex, data-rich systems while ensuring the desired level of utility is, however, challenging. 
What is considered an appropriate privacy/utility balance in one context might not be a proper trade-off for another context. 
Therefore, end-to-end system designs for privacy need to account for various privacy solutions and accommodate heterogeneous privacy preferences.
Next, we discuss some key design aspects for realizing end-to-end data privacy.
\fakeparagraph{User-Centric Privacy.}  
Users' perception of privacy varies widely across individuals, cultures, and contexts.
Therefore, the system needs to provide the means for users to set their privacy preferences and define how their data can be accessed, processed, and shared. 
In practice, user preferences can also vary with respect to the trade-offs between increased privacy and utility. 
Their preferences can vary based on the data involved and the target consumer. 
While we want to offer users the option of strong privacy guarantees, we also need to provide options for more relaxed privacy guarantees when incentives to do so exist.
For example, users might voluntarily share their off-platform shopping activities with a service provider in return for financial incentives~\cite{AmazonShoppingData}.
Therefore, a practical system needs to support a range of privacy preferences and be able to build privacy-compliant views across data covered by heterogeneous policies.

\renewcommand\cellalign{lt}
\newcommand*{\tabindent}{ \hspace{3mm}}
\begin{table}[]
	\centering
		\begin{threeparttable}[b]
			\resizebox{\columnwidth}{!}{%
			\begin{tabular}{lcl}
				\toprule
				\large{Name}                                                                          & \large{Zeph} & \large{Description}                            \\[2pt]
				\toprule
				\multicolumn{3}{c}{\textsc{\large{Data Masking}}} \\[2pt]
				Field Redaction~\cite{iridatamgmt, privitar,oracleredaction}                            & \CIRCLE     & Reveal some attributes and hide others        \\
				Predicate Redaction~\cite{iridatamgmt}                                                  & \LEFTcircle & Only reveal data that satisfy a predicate     \\
				\makecell{Det. Pseudonym. ~\cite{enisapseudonymization}}            & \Circle     & Replace value with a deterministic pseudonym  \\
				\makecell{Rand. Pseudonym.~\cite{enisapseudonymization}}           & \CIRCLE     & Replace value with a random pseudonym         \\
				Shifting~\cite{googledeidentify}                                                        & \CIRCLE     & Shift actual values by a fixed offset         \\
				Perturbation~\cite{Dwork2006-mp}                                                        & \CIRCLE     & \makecell{Perturb data by adding random noise \\[-1pt] i.e., additive differential privacy mechanism} \\
				\toprule
				\multicolumn{3}{c}{\textsc{\large{Data Generalization}}} \\[2pt]
				Bucketing~\cite{privitar,googledeidentify}                                              & \LEFTcircle & \makecell{Map values to a coarse space        \\[2pt]} \\
				Time Resolution~\cite{timecrypt}                                                             & \CIRCLE     & Aggregate data across time                    \\[2pt]
				Population~\cite{castelluccia2009, bonawitz2017practical, prio, honeycrisp-secure-aggr} & \CIRCLE     & \makecell{Aggregate data across a population  \\[2pt]} \\
				\bottomrule
			\end{tabular}
			}
			\caption{
				Overview of existing privacy transformations: Data Masking techniques (top), Data Generalization techniques (bottom).
					We use \CIRCLE~(full support),  
					\LEFTcircle~(partial
					support), and \Circle~(no support) to indicate which of these techniques are currently supported in \oursystem.}
			\ltab{tab:ptrans}
		\end{threeparttable}
\end{table}

\fakeparagraph{Retrofit into Existing Data Pipelines.}
A practical privacy solution should augment existing data pipelines
while ensuring the privacy of the underlying data.
Additionally, privacy transformations need to respect/adhere to traditional data protection mechanisms already in place (i.e., end-to-end encryption).
Therefore, the design needs to offer composability to support a variety of privacy solutions and ensure that privacy solutions can work with encrypted data.
We want to leave the flow of data in end-to-end encrypted systems intact.
\fakeparagraph{Privacy Transformations.} 
Privacy solutions for data analytics focus on allowing the use of data or computation on data subject to privacy restrictions specified by users (e.g., restrict what can be inferred from the data).
They are designed to enable extracting the utility from data while preserving individual's privacy preferences.
This is often achieved through a range of data modifications that we refer to as privacy transformations, i.e., functions applied to the data to limit and control the extent of sensitive information revealed to authorized parties.
Solutions in this space can be grouped into three broad classes~(\rtab{tab:ptrans}): \textit{(i)} data masking techniques that obfuscate sensitive parts of the data, 
\textit{(ii)} generalization techniques that reduce data fidelity, e.g., by aggregating data,  \textit{(iii)} combinations of  \textit{(i)} and  \textit{(ii)} which can realize complex transformations by chaining masking and generalization techniques. 
Privacy transformations are the primary tools to safely release data, achieving either a range of privacy guarantees common in practice (e.g., as in aggregate statistics) or formal privacy definitions such as k-anonymity~\cite{generalizationkanonimity} or differential privacy~\cite{dwork2006-dp}.
A useful end-to-end system design for privacy therefore needs to support a broad set of existing privacy transformations.

\setlength{\tabcolsep}{4pt}
\newcommand*\rot{\rotatebox{90}}

\begin{figure}[t]
	\center
	\includegraphics[width=1\columnwidth]{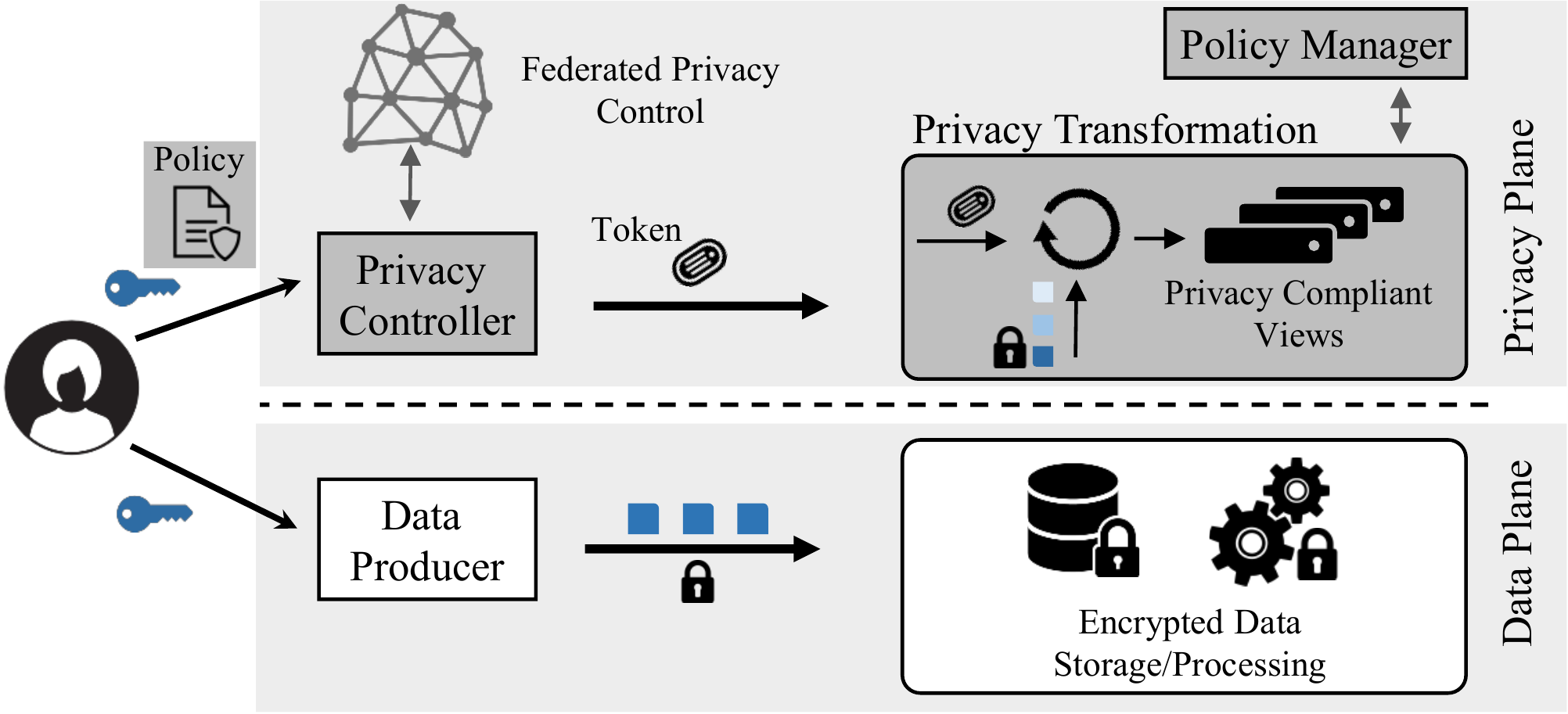}
	\caption{
	Overview of \oursystem's end-to-end approach to privacy. 
	}
	\lfig{overview}
\end{figure} 

\subsecspacingtop
\subsection{\oursystem in a Nutshell}
\subsecspacingbot

\oursystem is a privacy platform that augments encrypted stream processing pipelines with the means to enforce privacy controls cryptographically. 
\rfig{overview} shows an overview of \oursystem 's design.
We aim to enable authorized third-party services to access and process data and to gain insights from it without violating the privacy preferences of the data owners.
We design \oursystem to encapsulate state-of-the-art privacy solutions (e.g., generalization, differential privacy) while preserving the data flow in existing streaming pipelines.
\fakeparagraph{Privacy Plane.} To illustrate a deployment of \oursystem, we consider a health monitoring provider that stores health-related data from wearable devices such as heart rate and other metrics.
We assume that the data streams are already end-to-end encrypted, i.e., the wearables encrypt data before uploading, while applications (e.g., health dashboard) query encrypted data and locally decrypt the result~\cite{timecrypt}.
We refer to this data flow through a streaming platform as the \textit{data plane}. 
The privacy logic resides and is executed outside of the data plane, allowing \textit{data sources} to continue writing encrypted data streams to a remote stream processing pipeline as before.
\oursystem exposes an API for \emph{data owners} to set their privacy preferences, which forms the base for users' privacy policies.
A \oursystem deployment augments the data plane with a \textit{privacy plane} that enables the provider to extract information from the encrypted data streams through privacy transformers.
Hence, the service gets access to privacy-compliant transformed views of the underlying raw data streams.
For example, the service might collect the average heart-rate per day for different age-groups (i.e., population aggregate transformations).
To collect these statistics, \oursystem allows the service to express privacy options for data stream attributes through a modified data schema (\rsec{sec:privacypolicy}), which describes a set of possible privacy transformations for attributes.
Upon registering with the services, \emph{data owners} set their privacy preferences for each stream based on these options, forming the base for users' privacy policies.
For example, they can indicate if the service is allowed to include their heart rate stream in the specified aggregate transformations. 
The additional logic for handling privacy options and coordinating transformations is handled by an additional server component, the \comp{policy manager}.
The \comp{policy manager} offers an API to handle privacy options per data stream and coordinates privacy transformations as \comp{stream processors} in the streaming pipeline.
\fakeparagraph{Privacy Controller.} Policy enforcement in \oursystem is handled by the \comp{privacy controller}.
The privacy controller is responsible for supplying the cryptographic privacy transformation tokens that enable privacy transformations at the server.
As some privacy policies require data to be aggregated across different users before being made available, generating tokens specific to
these types of transformations require interaction between several privacy controllers in what we refer to as \emph{federated privacy control}.
While the tokens generated by the privacy controllers cryptographically enforce the data owners' privacy policies,
the server is responsible for composing and executing transformations efficiently.
The \comp{privacy controller} does not require access to the data and can be hosted in a location with higher availability guarantees.
\oursystem allows users to choose a variety of deployment scenarios for privacy controllers. 
Privacy controllers could be self-hosted, hosted on-premise for corporations, or outsourced to a trusted provider (e.g., OpenID identity providers).
\fakeparagraph{Data Consumers.} We distinguish between two types of \emph{data consumers}: (a)~services that access the data to provide utility to the user (i.e., personalization), and (b)~third-party services, e.g., to provide a utility that is beneficial to the public or the service itself, but not directly to the user (e.g., allow your health data records to contribute to a medical study).
Enabling direct access to the data for the first type of data consumers is handled by cryptographic access control and is supported in our design, but it is not the focus of this work.
In \oursystem, we focus instead on the latter with the goal to continue enabling the benefits of these services while respecting users' privacy.

\subsecspacingtop
\subsection{Threat Model}
\subsecspacingbot
\label{threat}

\oursystem enforces users' privacy preferences cryptographically, i.e., users are guaranteed that the data is transformed with the privacy transformation corresponding to their privacy preference before it is released to applications. 
Meanwhile, their original data remains end-to-end encrypted.      
\fakeparagraph{Setting.} 
We assume an \emph{honest-but-curious}~\cite{cryptdb} server, i.e., the server performs the computations correctly but will analyze all observed data to gain as much information as possible.
We also assume the existence of a public-key infrastructure (PKI) for authentication of privacy controllers/data producers. 
In this setting, \oursystem ensures \emph{data confidentiality}, more specifically \emph{input privacy}, guaranteeing that the adversary learns nothing about the raw data streams except what can be learned from the output of the transformation $F$ (i.e., $\hat{F}$\textit{-privacy}~\cite{prio}) with some modest leakage function due to encodings (\rsec{sec:encryption}). 
\oursystem also ensures that an adversary controlling the server and at most a fraction $\alpha$ of privacy controllers 
is unable to violate the privacy policies of other data owners.
\fakeparagraph{Data Plane.}
In \oursystem, data streams are encrypted at the source with a semantically secure encryption scheme, while the metadata (e.g., timestamps) is sent in plaintext.
Decryption keys are never disclosed to the server; therefore, raw data confidentiality is guaranteed even in the case of a server compromise.
If an adversary gains control over a data producer or the responsible privacy controller, only the data associated with that producer/controller is revealed.
\fakeparagraph{Privacy Plane.}
\oursystem ensures input privacy for honest data owners even if the stream processor executing a privacy transformation or the policy manager coordinating it is compromised by an adversary.
For the case where $F$ is an aggregation function involving data from different privacy controllers (i.e., federated privacy control), we assume that at most a fraction of $\alpha$ of the entities in the aggregation transformation are controlled by the adversary.
Note that this can also include the server.
The choice of $\alpha$ depends on the deployment scenario. In ~(\rsec{sec:secure-aggregation}), we show how this choice affects performance.
For our evaluation, we use a pessimistic value of $\alpha = 0.5$, but real-world deployments might use significantly lower values.

\fakeparagraph{Robustness.}
While \oursystem can handle various failures in practice, formal robustness against misconfigured or malicious privacy controllers or data producers is out of scope for this design. 
A privacy controller sending corrupted tokens cannot compromise privacy but could alter the output of a transformation or prevent a transformation from completing.

\vspace{-5pt}
\section{Encryption for Privacy Transformations}
\lsec{sec:encryption}

In this section, we describe our approach to enable privacy transformations in end-to-end encrypted 
systems. 
Our design serves privacy-compliant transformed views of data without affecting the data flow of an end-
to-end encrypted stream-processing system.
To meet this goal, our design \mbox{logically decouples} privacy transformations and policy enforcement from the generation and storage of data.
Data producers remain oblivious to the transformations and do not need to encrypt data towards a fixed privacy policy.
The modifications needed for privacy transformations are instead executed outside the data plane (i.e., conventional data flow), 
working exclusively on encryption keys to generate what we call cryptographic \emph{transformation tokens}.
These tokens are, in essence, the necessary cryptographic keying material that enables the respective transformation on encrypted data.
Outputs of privacy transformations over encrypted data at the server-side are then released by combining the encrypted data with corresponding cryptographic transformation tokens.
Introducing a logical separation between the data plane and privacy plane %
allows for heterogeneous policies atop the same data and leaves the conventional data flow unaffected.
In this realization of \oursystem we focus on streaming data. Hence, we focus on cryptographic building blocks that optimize efficiency for this type of data.
The design concepts underpinning \oursystem are generic and could be adapted to other systems using other cryptographic building blocks. 
These, however, can introduce their own trade-offs between computation \mbox{expressiveness and performance}.

\vspace{-10pt}
\subsecspacingtop
\subsection{Decoupling Encryption from Privacy \\ Transformations}
\subsecspacingbot

This design requires an encryption approach that supports \textit{homomorphic evaluation} in a variety of settings.
Namely, it needs to support: \textit{(i)}~evaluation on encrypted data, i.e., encrypted data processing, 
\textit{(ii)}~construction of cryptographic \emph{transformation tokens}, and
\textit{(iii)}~combining the encrypted data with the matching cryptographic transformation tokens for selective release of privacy-compliant transformed data views.

\fakeparagraph{Encrypted Data Processing.}
Existing encrypted data processing systems utilize homomorphic encryption schemes to enable server-side computation on encrypted data~\cite{timecrypt,seabed,cryptdb,monomi}.
To meet applications' stringent performance requirements,
systems typically combine efficient partially homomorphic encryption schemes~\cite{timecrypt,seabed,prio,ml_homenc_largescale,ml_encoding}
with specialized client-side encodings to support a wider set of queries.
However, standard homomorphic encryption schemes do not lend themselves to support selective release of data (i.e., handing out the decryption key allows to decrypt all data) 
or support privacy transformations that require data evaluation across different users (i.e., different trust domains).
Supporting functions across populations in the multi-client/single-server setting is typically achieved via specialized multi-party computation protocols~\cite{prio,sketch,castelluccia2011CancelOut,honeycrisp-secure-aggr}. %
These existing protocols ensure that the user inputs remain private and only the output of the function evaluation is revealed to the server.
However, they require active participation by the data producers and are often limited to specific functions.
We want to remove the need for -- potentially resource-limited -- data producers to take part in or even be aware of privacy transformations.

\fakeparagraph{Homomorphic Secret Sharing.} 
To decouple encryption from privacy transformations, we draw on ideas from the Homomorphic Secret Sharing (HSS)~\cite{hom-secret-share1, hom-secret-share2} literature.
In essence, HSS allows computing a function $F$ on secret shared messages by combining the outputs of a function $\hat{F}$ applied on the individual secret shares.
HSS could be used to split stream events into two shares: one for the data plane (server) and one for the privacy plane.
The privacy plane could authorize a transformation $F$ by computing the same function $\hat{F}$ as the server on their local input shares, and releasing the output.
Here, $\hat{F}$ supports all of the required core functions, as secret shares can also be aggregated across different data owners.
Applying standard HSS in our setting raises two issues: \textit{(i)}~general-purpose HSS incurs non-negligible overhead~\cite{hom-secret-share1}, and \textit{(ii)}~with this approach privacy controllers remain dependent on data producers as they continue to receive a secret share for each new stream event.

To address the first issue, we employ \emph{additively} homomorphic secret sharing.
    This is considerably more efficient than general-purpose HSS, allowing our system to sustain the high throughput needed for streaming data workloads.
    Used naively, additively homomorphic secret sharing can significantly limit expressiveness.  
    However, as we show in the next section, using carefully selected data encodings allows us to support a wide set of privacy transformations.

To break the dependency between privacy controllers and data producers,
we enable the privacy controller to independently derive the tokens (i.e., its shares) based only on metadata about the stream.
Given a shared common master secret, the data producer and privacy controller never have to communicate or even be online at the same time.
We introduce our scheme in more detail in \rsec{sec:tokens}.
Our tailored scheme offers both the required efficiency and the flexibility necessary to decouple the data plane from the privacy plane.
The data producer and the responsible privacy controller need to only agree on a shared master secret.
Then, the privacy plane can authorize a transformation $F$ by deriving the involved shares and executing $\hat{F}$ on them, which results in a \textit{transformation token}.
This token allows the server to compute and reveal the output of $F$ by performing $\hat{F}$ on the ciphertexts and combining the result with the \textit{transformation token}.
If the transformation $F$ spans multiple trust domains, i.e., the privacy plane consists of multiple privacy controllers,
the privacy controllers run an MPC protocol to compute the final transformation token.
Note that this does not require the data producers to participate or even be online.
Next, we show how we can support a broad set of privacy transformations with this scheme.

\subsecspacingtop
\subsection{Privacy Transformation Functions}
\subsecspacingbot
\lsec{tranformation-functions}

Broadly, privacy transformations (~\rsec{privacy-transformations}) generally involve computation/perturbation of individual user's data, computations across different users' data, or combinations \mbox{of the two}.  
Based on this insight, 
\oursystem exposes three core functions for developers that allow for privacy transformations in the encrypted setting:
\textit{(i)}~\funcs, which enable ciphertext aggregation operations within the same user's data streams.
\textit{(ii)}~\funcm, which enables ciphertext aggregation across streams from a population of users,
\textit{(iii)}~\funcdp, which supports perturbation via noise addition to streams aggregated across multiple users.

\fakeparagraph{Privacy Transformations in \oursystem.}
A privacy transformation $F$ in \oursystem is realized by combining a chain of core functions
and/or withholding certain shares when creating a token.
\textit{\textit{(i)}~Data Masking.}
Data masking techniques such as field-redaction and randomized pseudonymization are directly supported by the secrecy properties of our scheme.
The privacy controller redacts or pseudonymizes a field by withholding the corresponding shares from the transformation token. 
Shifting and perturbation are realized with \funcs by adding a constant or calibrated random offset to the transformation token.
\oursystem supports a subset of predicate redactions using client-side encodings that represent a value as a vector of elements.
A privacy controller can then construct a transformation token that only reveals a subset of elements in the vector or a certain sum of the elements (\funcs).
For example, to enable predicate redaction based on a threshold, the client would encode the value as a vector of two elements. 
If the value is above the threshold, the client stores the value as the first element in the vector or else as the second element.
To only reveal the values above the threshold, the privacy controller can disclose the first elements of the vectors with the tokens.

\textit{\textit{(ii)}~Data Generalization.}
Bucketing similarly builds on client-side encodings that map a value to a one-hot vector representing the whole domain. 
Instead of releasing a token for all elements, the privacy controller uses \funcs to release a sum of shares for elements mapping to the same bucket.
For values with a large domain, we can approximate the frequency count with a histogram using a larger bin width.
\oursystem supports data generalization over time with \funcs and population with \funcm. 
Moreover, \oursystem provides \funcdp to release a differentially private aggregate across a population.
To extend the supported aggregation functions, we leverage existing encoding techniques~\cite{timecrypt,seabed,prio,ml_homenc_largescale,ml_encoding, encodings-splinter}.
In essence, these encodings map a value to a vector with different statistics that allows the computing platform to execute functions by performing element-wise addition.
The aggregate functions sum and count are inherently additions. 
With a vector of sum and count, a party can obtain the mean by dividing the sum by the count.
By adding the square of a value to the encoding vector, a party can calculate the variance using that $Var(x) = \mathbb{E}(x^2)-\mathbb{E}(x)^2$.
Moreover, with the one-hot encoding, constructing a histogram corresponds to the element-wise sum of a set of one-hot vectors. Given a histogram, a party can compute the median or other percentiles, min, max, mode, range, or topk.
Prior work~\cite{prio} presents further encoding techniques for other functions that we support in \oursystem.

\subsecspacingtop
\subsection{Transformation Tokens}
\subsecspacingbot
\lsec{sec:tokens}

In \oursystem, we build upon a symmetric homomorphic encryption scheme~\cite{timecrypt} explicitly designed for streaming workloads.
We use this scheme to realize efficient additively homomorphic secret sharing for our setting.
The scheme efficiently derives a unique sub-key for each message from a master secret
and encryption is performed via modular addition of the key and the message.
Here, the encrypted message and the (message-specific) sub-key can be seen as additive shares of the message.
Since encryption and decryption are linear operations, the scheme supports linear aggregation by computing the function on both the sub-keys and the encrypted messages independently.
\textit{Transformation tokens}, which authorize the release of privacy transformation results, are derived from the sub-keys via aggregations.
We now describe how these are constructed in our system.
We start with a description of a simplified version of \oursystem that assumes a single privacy controller and extend our description to consider multiple \mbox{privacy controllers in}~\rsec{sec:secure-aggregation}.

\fakeparagraph{Symmetric Homomorphic Stream Encryption.}
First, we give a brief summary of the symmetric homomorphic stream encryption scheme~\cite{timecrypt} we build upon.
Let a data stream be a continuous stream of events $\{e_0, e_1,...,e_i, e_{i+1},...\}$ for events $e_i := (t_i, m_i)$ consisting of a message and a timestamp.
Each message $m_i$ is an integer modulo $M$ and is annotated with a discrete timestamp $t_i \in I$.
We assume events are ordered by their timestamps and created in-order.

In the setup phase, a master secret $k$ is generated, the group size $M$ is defined (e.g., $2^{64}$), and a keyed pseudo-random function (PRF) $f_k: I \rightarrow [0, M-1]$ that outputs a fresh pseudo-random key $k_i$ for timestamp $t_i$ is selected.
To encrypt an event $e_i$, the data producer uses the event timestamp from the last encrypted message $t_{i-1}$ and computes:
\begin{equation}
Enc(k, t_{i-1}, e_i) = (t_i, t_{i-1}, m_i + k_i - k_{i-1} \mod M)
\end{equation}
where ${k_i =  f_k(t_i)}$, $k_{i-1} = f_k(t_{i-1})$.
This scheme is additively homomorphic: ciphertexts can be aggregated via modular additions.
Keys can be aggregated the same way, but for a time-window $[t_i, t_j]$,
the client can decrypt more efficiently
by deriving only the two outer keys ${k_{i} =  f_k(t_{i})}$ and ${k_{j} =  f_k(t_{j})}$ because the inner keys cancel out.
This encryption scheme hides the inputs from the server and allows the server to perform aggregations among the streams without accessing the plaintext data.
\fakeparagraph{Authorizing Transformations.}
The intuition in \oursystem is that the encryption scheme essentially splits a message $m_i$ %
into two additive secret shares: the key $- k_i + k_{i-1}$ and the ciphertext $c_i$, with  $m_i = c_i + (- k_i + k_{i-1}) \mod M$.
Therefore, any transformation $F$ consisting of the three core aggregate operations
can be performed independently on both the ciphertexts and the keys using modular additions.
The latter produces a \textit{transformation token} %
$\tau_F$, which the server can use to reveal the output $o_F$ of a transformation $F$ by \mbox{computing $o_F + \tau_F \mod M$}.
Hence, a privacy controller that is in possession of the master keys of the streams can authorize a transformation $F$ by deriving the necessary keys and performing the transformation on top of them to produce a matching \textit{transformation token} $\tau_F$.
In the following, we assume that all additions are performed modulo the parameter $M$.

\fakeparagraph{Single-Stream Transformation Tokens.}
We now describe how a privacy controller can create transformation tokens for \funcs transformations, e.g., only reveal the approximate locations aggregated over a month.
We start with a window aggregation to reduce the time resolution. The server adds values within the specified time window $t_{i}$ to $t_{i + w}$ , where $w$ is the window size.
As long as data producers submit a value on each window border, the resulting ciphertext of the window aggregation on the server shares has the form $c_w = m_{aggr} + k_{i + w} - k_{i - 1}$.
The privacy controller can compute the transformation token for this window $\tau = - k_{i + w} + k_{i - 1}$ by deriving only the two outer keys ${k_{i} =  f_k(t_{i})}$ and ${k_{j} =  f_k(t_{j})}$ as the inner-keys cancel each other out~\cite{timecrypt, seabed}.
With this token, the server can decrypt the window aggregation if and only if the correct windows were aggregated, as the keys directly encode the window range.
For aggregations within events, the privacy controller uses modular addition to add the respective sub-keys to create the transformation token.
The privacy controller can construct transformation tokens for values with encodings (\rsec{tranformation-functions}) by selectively releasing the sub-keys of certain elements in the encoding vector or by aggregating sub-keys of elements in the vector.

\fakeparagraph{Multi-Stream Transformation Tokens.}
Multi-stream transformation tokens reveal the output of \funcm transformations, which aggregate data over multiple streams, e.g., only reveal the approximate location aggregated among multiple users.
In multi-stream aggregation, the server sums a fixed window  $t_{i}$ to $t_{i + w}$ across different streams.
Let $S$ be the set of streams in the aggregation.
For each stream $j \in S$ we have a window aggregated share
$c_w^{(j)} = m_{aggr}^{(j)}  + k_{aggr}^{(j)}$ where $k_{aggr}^{(j)} = k_{i + w}^{(j)}  - k_{i - 1}^{(j)}$.
The aggregation over all streams in $S$ results in the sum of all window aggregates and the sum of all window share keys.
Hence, a privacy controller can compute the transformation token by aggregating the window keys  $\tau^{(j)} = -\sum_{j \in S} k_{aggr}^{(j)}$.

\fakeparagraph{Differentially-Private Transformations.}
\lsec{sec:noise}
Differential Privacy~\cite{dwork2006-dp} provides formal bounds on the leakage of an individual's private information in aggregate statistics.
The most common technique to achieve a differentially private release of information is to add carefully calibrated noise.
\oursystem supports \emph{noisy} transformations (i.e., \funcdp) on multi-stream window transformations, but could be extended to the single-stream setting.
The privacy controllers add carefully calibrated noise to the keys (i.e., submit noisy keys): $\tilde{\tau_j} = \tau_j + \eta_j$ where $\eta_j$ is the noise.
\oursystem therefore supports all additive noise mechanisms from the Differential Privacy literature~\cite{Dwork2006-mp} with noise drawn from a divisible distribution.
However, mechanisms like the \emph{Sparse Vector Technique}~\cite{privacybook} that require access to the underlying data cannot be applied this way.
In previous work, noise is added to plaintexts prior to encryption, whereas in \oursystem noise is added to the decryption keys.
The two approaches are cryptographically equivalent. 
However, previous work requires deciding on the noise to add at encryption time.
Our approach has the advantage of allowing noise to be added to data that was previously encrypted without consideration for noise.
This also means, that the same data is reusable for encrypted storage and to facilitate one or multiple differentially private privacy transformations.

\subsecspacingtop
\subsection{Transformations Across Different \\Trust Domains}
\subsecspacingbot
\lsec{sec:secure-aggregation}
Until now we assumed a single privacy controller that is in control of all streams.
We now discuss how \oursystem enables multiple privacy controllers that are each responsible for a distinct subset of streams.
While we assume that data owners trust their own privacy controller,
different data owners might not want to trust the \emph{same} controller. %
In such multi-trust setting, the server needs to interact with all privacy controllers involved in a transformation.
Hence, when aggregating across streams the server needs to request a transformation token from each privacy controller.
In a na\"ive approach, the privacy controllers might simply send a combined token for the aggregation of the streams under their control.
However, this leaks the intermediate result from each controller to the server.
Instead, we need the individual tokens to reveal no additional information while still enabling correct decryption of the transformation output.
We enable this in \oursystem using secure aggregation~\cite{castelluccia2011CancelOut, bonawitz2017practical}, a specialized secure MPC protocol.
For our system, we require a secure aggregation protocol that is \textit{(i)} lightweight in terms of computation for privacy controllers and~\textit{(ii)} can be efficiently executed multiple times with similar participants.
Based on these requirements, \oursystem builds on the secure aggregation protocol from Ács et al.~\cite{castelluccia2011CancelOut} to create transformation tokens over multiple parties.
The protocol goes hand in hand with the design of the transformation tokens, as it also relies on additive masking. %
In the following, we outline the core protocol and then describe our optimizations that reduce the computation cost \mbox{for privacy controllers}.
\fakeparagraph{Core Protocol.}
We consider a set $\mathcal{P}$ consisting of $N$ privacy controllers and a server that aggregates the inputs.
Each privacy controller $p \in \mathcal{P}$ owns a token $\tau_p$ that is constructed by aggregating the tokens for the corresponding \funcs transformation for each stream under their control.
The goal of the protocol is to compute $\tau = - \sum_{p \in \mathcal{P}} \tau_p$ without revealing the individual inputs $\tau_p$ to the server or to the other privacy controllers.
Each privacy controller masks its input $\tau_p$ with a nonce $k_p$, i.e., it computes $\tau_p + k_p \mod M$.
The nonces are constructed such that the sum over all nonces results in $\sum_{p \in \mathcal{P}} k_p = 0$.
As a consequence, the \mbox{sum over all encrypted inputs results in the sum of inputs}:

\begin{equation}
	\sum_{p \in \mathcal{P}} \tau_p + \sum_{p \in \mathcal{P}} k_p \mod M = \sum_{p \in \mathcal{P}} \tau_p \mod M
\end{equation}

To construct the canceling nonce, each privacy controller establishes $N-1$ pairwise shared secrets $k'_{p,q}$ with all other privacy controllers
which are aggregated to form the nonce $k_p$.
In particular, if $p > q$, then the controller $p$ adds $-k'_{p,q}$ else $k'_{p,q}$.
\begin{equation}
k_p = \sum_{p > q} -k'_{p,q} + \sum_{p < q} k'_{p,q} \mod M
\end{equation}
Hence, the pairwise secrets cancel each other out when the masks are combined in the aggregation.
For conciseness, we refer to Ács et al.~\cite{castelluccia2011CancelOut} for a description of dropout handling.

\fakeparagraph{Constructing Canceling Nonces.}
In \oursystem, the secure aggregation protocol is run repeatedly for multiple rounds due to the continuous nature of streaming queries.
Thus, privacy controllers require an efficient method to establish many pairwise shared secrets.
The standard protocol achieves this with a setup phase where the parties create pairwise shared secrets $k_{p,q}$ using a Diffie-Hellman key exchange.
These pairwise secrets then serve as seeds (or keys) for a PRF to establish nonces for each round $r$: $k_{p,q}^{r} = PRF(k_{p,q}, r)$.
Even though PRF computations are significantly more efficient than a Diffie-Hellman key exchange, this protocol still requires each privacy controller to evaluate $O(N)$ PRF's and additions to create the blinding nonce $k_p$ for a single transformation token,
which can be expensive for large $N$.

To improve this theoretical overhead, we view the complexity of creating a shared blinding nonce as a graph $G=(V,E)$ with the 
set of vertices $V$ representing the involved parties ($|V| = N$), and the set of edges $E$ denoting the pairwise canceling 
masks $k'_{p,q}$.
In the standard form described above, the graph $G$ forms a Clique because every privacy controller includes a pairwise mask 
$k'_{p,q}$ with every other privacy controller.
To reduce the number of PRF evaluations in the online phase for a privacy controller (i.e., reduce the number of edges in the 
graph $G$), we propose an optimization that leverages the fact that the protocol is repeated  over a long period of time with 
similar participants, i.e., the long-running \mbox{nature of streaming queries}.

\fakeparagraph{Online Phase Optimization.}
We reduce the communication overhead during the online phase
     by choosing privacy controller's nonce as to only include a small random subset of the pairwise-secrets in each round.
In graph terms, this corresponds to a small expected degree of each vertex.
As long as the graph remains connected\footnote{More specifically, the subgraph of \emph{honest} nodes must remain connected.}, confidentiality is guaranteed.
We divide the online phase into epochs consisting of $t$ rounds.
At the beginning of each epoch, we use $N-1$ evaluations of the PRF to bootstrap the secure aggregation graphs for the epoch.
A privacy controller assigns each edge to a small number of rounds, based on the output of a PRF evaluated on the shared secrets.
More specifically, we divide the output of $PRF(k_{p,q}, r)$, where $r$ is a public epoch-identifier, into $b$-bit segments.
Each segment assigns the edge $e_{p,q}$ to one of $2^b$ graphs using the number encoded in the $b$-bit segment.

Assuming a 128-bit output size of a PRF (e.g., AES), an epoch consists of $t = \lfloor 128/b \rfloor \cdot 2^b$ rounds.
In comparison, the protocol of Ács et al.~\cite{castelluccia2011CancelOut} uses the same $N-1$ PRF evaluations to create only a single secure aggregation graph (i.e., epoch size of one).
Ideally, we want to create as many graphs as possible, i.e., select a large $b$, since with increasing $b$, an epoch consists of more rounds.
However, with increasing $b$, each of the associated graphs has fewer edges%
, which increases the risk of a graph being disconnected. %
In the appendix~\rsec{secaggop:detail}, we show how to select $b$ so that the probability of any honest subset of nodes being isolated in any of the $t$ generated graphs is bounded by $\delta$, assuming a fraction of at most $\alpha$ parties collude.

For example, for 10k privacy controllers, assuming that up to half are colluding ($\alpha = 0.5$), and bounding the failure probability by $\delta = 1 \times 10^{-9}$, allows for $b=7$, which results in an epoch consisting of 2304 rounds where each vertex has a expected degree of 78.
As a consequence, our optimization requires 190k PRF evaluations and 180k additions for constructing all 2304 blinding nonces of an epoch.
In comparison, the basic protocol requires 23 million PRF evaluations and additions while the protocol from Ács et al.~\cite{castelluccia2011CancelOut} requires 23.2 million PRF evaluations and 180k additions\footnote{All results assume that $\tau_p$ is at most 128-bit long and hence a single evaluation of AES is sufficient for encryption.}.

\secspacingtop
\section{\oursystem System Design}
\secspacingbot
\lsec{design}

\begin{figure}[t]
	\center
	\includegraphics[width=1\columnwidth]{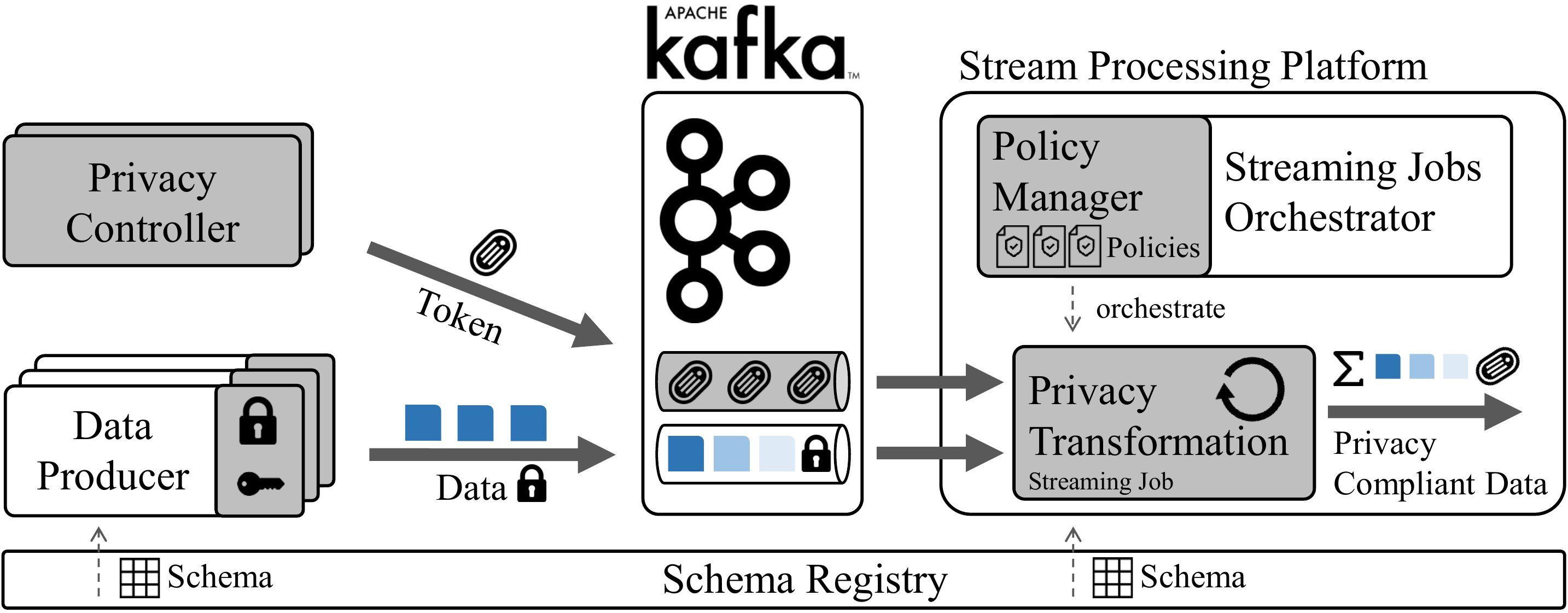}
	\caption{
    Overview of \oursystem's architecture and integration into existing data streaming pipelines. \oursystem's components are highlighted in gray.}
    
	\lfig{arch}
\end{figure}

\oursystem is a privacy platform that cryptographically enforces user-defined privacy preferences in streaming platforms by sharing only transformed privacy-compliant views of the underlying encrypted data.
So far, we have described the cryptographic building blocks that enable privacy transformations in \oursystem.
Here, we describe how we overcome the system challenges that need to be addressed to allow practical deployment.
\oursystem augments existing stream processing pipelines, similar to existing frameworks operating on data in-the-clear~\cite{privitar-kafka}:
\textit{(i)} On data producers, \oursystem adds a proxy module for encoding and encryption.
\textit{(ii)} On the server, \oursystem adds a microservice running in the existing stream processing platform. 
This microservice transforms the incoming encrypted streams into privacy-compliant output streams (\rfig{arch}),
which can then be consumed by existing stream processing queries for arbitrary post-processing.

\subsecspacingtop
\subsection{User API and Privacy Policies}
\subsecspacingbot
\lsec{sec:privacypolicy}

Before introducing the \oursystem components in detail, we discuss aspects related to users' interaction with \oursystem.

\fakeparagraph{Privacy Preferences.} 
\oursystem provides the capabilities for users to set their privacy preferences (i.e., user-centric privacy)
and the means to cryptographically enforce various privacy policies in a unified system.
In this paper, we do not consider the question of what this set of privacy preferences should be.
Nevertheless, we suggest and implement a sensible set of options to demonstrate how \oursystem can be used in practice. %
In the current design, data owners can set their preferences as follows: \textit{(i)} do not share my data, \textit{(ii)} share my data without restrictions, \textit{(iii)} share my data only when aggregated with other users, and  \textit{(iv)} 
share only generalized views of my data and/or mask sensitive data, i.e., share but limit inference of sensitive information from my data.
The realization of these preferences in practice is application- and data-dependent (i.e., generalization and data minimization techniques can differ depending on the data type, e.g., image, location, heart rate).

\fakeparagraph{Data Stream Schema.}
In \oursystem, developers can translate user preferences to an application-specific set of transformations by mapping them in a schema language.
\oursystem's schema language builds on the \emph{Avro}~\cite{avro} schema language (\rfig{fig:schema-and-policy}).
Using our extended schema language, developers can translate users' privacy options to configurations, encodings, and transformations for their application.
In addition, the schema contains meta-information about the stream and the contents of events within a stream.
This enables seamless integration into existing streaming services employing schema registries to store structural information about the events flowing through the system.
A \oursystem stream schema contains:
\textit{(i) Metadata attributes} describing static fields that remain constant for an extended period of time and are public information.
\oursystem's microservice uses these metadata tags to group and filter streams for transformations over different populations 
(\rsec{sec:query-matching}).
For example, the region where a data stream originates from (\rfig{fig:schema-and-policy}).
\textit{(ii) Stream attributes} describe the private contents of an event message and are annotated with all possible supported queries.
These explicit annotations are required to derive the necessary encodings to 
execute queries using the three core functions (\rsec{sec:encryption}).
For example, a heart rate sensor might have two stream attributes such as heart-rate and heart-rate variability (\rfig{fig:schema-and-policy}).
The heart-rate is annotated to support aggregates with variance statistics.
\textit{(iii)} The \textit{privacy options} for stream attributes.
A privacy option describes the set of transformations that the service can perform to reveal an output.
The options \emph{stream-aggregate}~(\funcs), \emph{aggregate}~(\funcm), and \emph{dp-aggregate}~(\funcdp) directly correspond to the three core functions defined in \rsec{sec:encryption}.
In addition,
\emph{private} does not allow any transformations on the stream while \emph{public} allows access to the raw data.
For each transformation set, one can add further constraints, e.g., defining a minimum population size, specifying a lower temporal resolution by aggregating over time or providing a privacy budget for the transformation.

\fakeparagraph{Annotating Streams.}
The \oursystem schema for a particular application can be accessed by all privacy controllers.
A user's privacy selection in the application triggers the responsible privacy controller to create a matching stream annotation and share it with the server.
A stream annotation contains the selected privacy option along with values of the metadata attributes and additional information about the stream (\rfig{fig:schema-and-policy}).
This information later allows \oursystem's server to identify suitable streams to include in privacy transformations.
Stream annotations contain an identifier of the data owner (e.g., the hash of their public key) that maps to the data owner's public key in the PKI.

\begin{figure}[t]
	\center
  \includegraphics[width=\columnwidth]{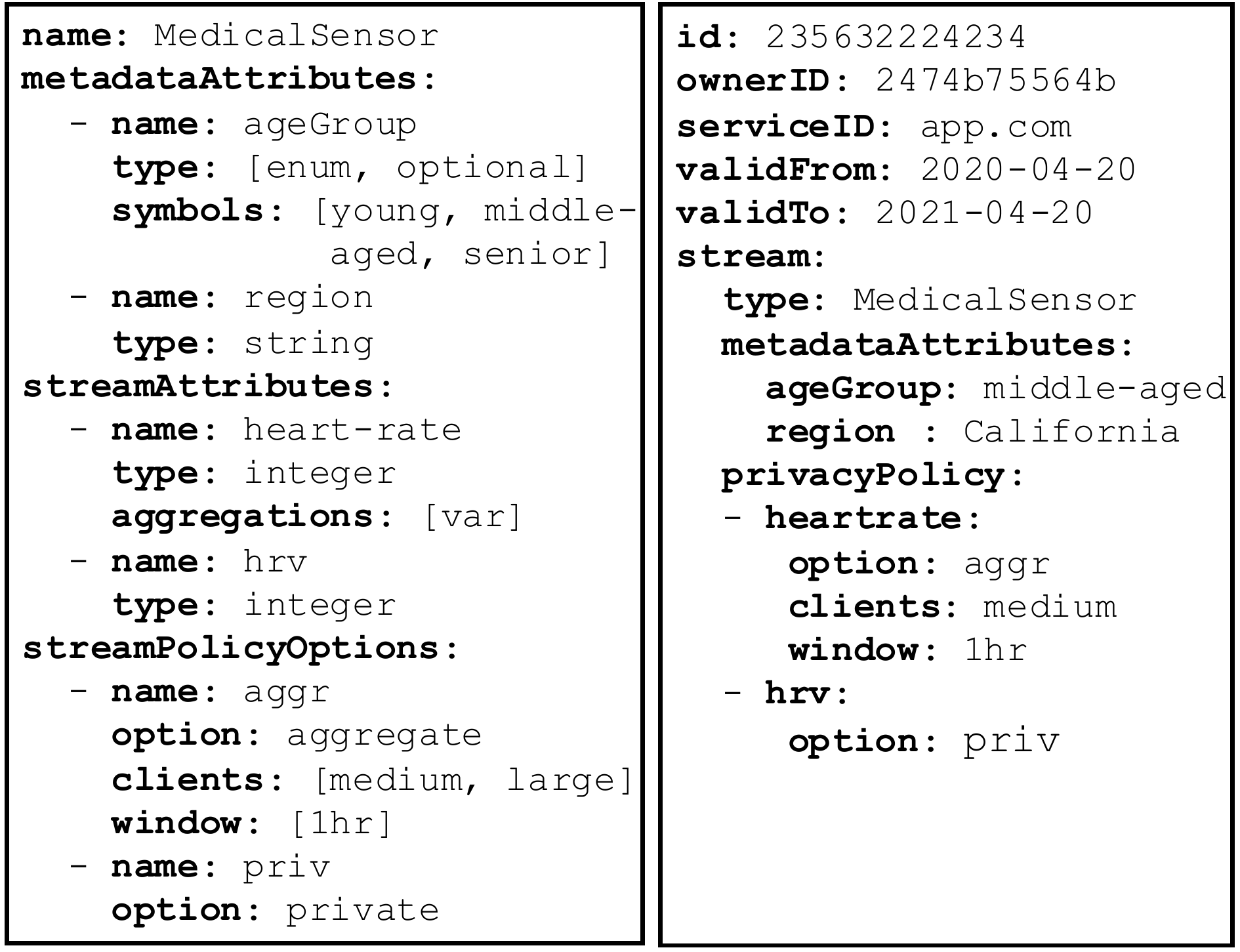}
	\caption{
    An example privacy policy schema of a medical sensor (\textit{left}) and a stream annotation for this schema (\textit{right}).
		(YAML format for display)
	}
  \lfig{fig:schema-and-policy}
\end{figure}

\vspace{10pt}
\subsecspacingtop
\subsection{Writing Encrypted Data Streams}
\subsecspacingbot
\lsec{sec:design:writing-data}
Data producers submit streams of events to the pipeline where each event conforms to a data schema in the schema registry, as in standard streaming pipelines.
However, \oursystem augments data producers with a proxy module to handle \mbox{encoding and encryption}.

\fakeparagraph{Setup.}
To initialize a new data stream matching a \oursystem schema, the data producer generates a master secret and shares both the schema and the master secret with the associated privacy controller.
After the initial setup phase, the data producer can start sending encrypted data to the server without any further coordination with the privacy controller.

\fakeparagraph{Encrypting Data Streams.}
The proxy module encrypts each record with a symmetric homomorphic encryption scheme~(\rsec{sec:encryption}), using the master secret from the setup phase.
\mbox{In order to} allow the privacy controller to derive a transformation token without observing the data (\rsec{sec:encryption}),
    the data producer sends a neutral value at regular intervals (e.g., every minute) to terminate the window.
    This does not affect the result of computations but is required for efficient \funcs transformations across time.
Additionally, these messages allow \oursystem's microservice to detect and handle dropout of data producers (e.g., due to network interruptions).

\subsecspacingtop
\subsection{Matching Queries with Privacy Policies}
\subsecspacingbot
\lsec{sec:query-matching}
\oursystem's microservice contains a policy manager that maintains a global view of the system and coordinates active streams, privacy controllers, and transformations in the streaming pipeline.
It provides a query interface for launching new transformations and matches queries with available streams by considering their chosen privacy options.
Privacy transformations are constructed from chains of the core operations~(\rsec{tranformation-functions}) 
and are executed as stream processing queries running continuously on a set of encrypted streams.

\oursystem's policy manager includes a query planner that leverages the fact that privacy transformation queries follow the same structure, which we discuss in more detail below.
The policy manager needs to ensure that queries comply with all the stream's selected policy options. Otherwise, it will not receive the required transformation tokens from the privacy controllers. %

\fakeparagraph{Query Language.}
The query language of \oursystem builds on \emph{ksql} \cite{ksql-jafarpour2019}, an SQL-like query language for expressing continuous queries on data streams.
Any authorized service can express privacy transformations that follow the pattern explained above.
\rfig{fig:query-planner} shows an example query, which creates a transformed stream for the hourly average heart rate of seniors in California, including at most 1k streams.

\fakeparagraph{Query Planner.}
The query planner executes queries from authorized services in three steps:
\textit{(i)} streams are filtered by their metadata attributes (e.g., all medical sensor streams in California).
\textit{(ii)} an \funcs operation using a time-window is performed on certain attributes of each selected stream (e.g., average heart rate over 1 hour).
The query planner checks for each selected stream that the transformation complies with the annotated privacy options for the attributes used, else the stream is excluded.
\textit{(iii)} If more than one stream is selected, an \funcm or \funcdp operation is performed on the results of the previous step.
The query planner checks for each remaining stream that the transformation complies with the \mbox{privacy option} and checks that the population constraints are met (e.g., minimum population size), or otherwise excludes the stream.
These compliance checks are necessary, as privacy controllers would not provide the required tokens for a stream where the privacy options do not allow the query.
To prevent an attacker from combining outputs of different transformations to violate privacy policies, any stream attribute can be matched to only one transformation, and is removed from the set of queriable streams for this attribute as long as the stream is part of the running transformation. %
The privacy controller generally only supplies a single transformation token for each window in a given stream, preventing differencing or re-use attacks. 
For DP aggregations, a stream value can contribute to multiple transformations if allowed by its current privacy budget. 
The privacy controller maintains the privacy budget and suppresses transformation tokens if the privacy budget is used up.
After processing the query, the query planner outputs a \textit{transformation plan} that encodes the list of streams in the transformation, fault tolerance details (i.e., number of participants dropout the system can handle), and the sequence of operations the system need to perform (\rfig{fig:query-planner}).

\begin{figure}[t]
	\center
	\includegraphics[width=\columnwidth]{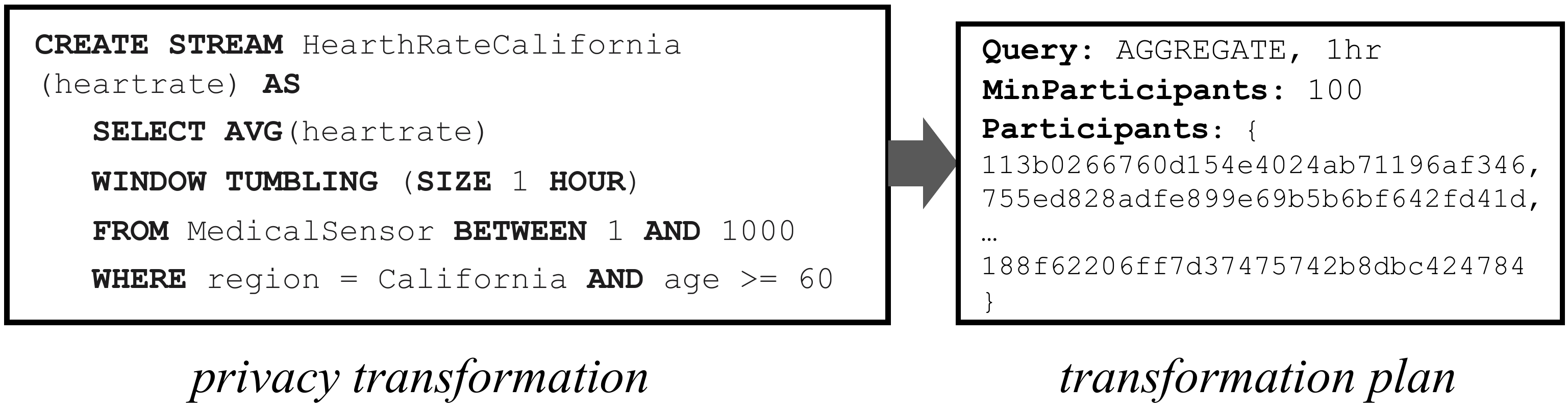}
	\caption{
		The query planner converts privacy transformations into transformation plans with complying data streams.
	}
  \lfig{fig:query-planner}
\end{figure}

\subsecspacingtop
\subsection{Coordinating Privacy Transformations}
\subsecspacingbot
\lsec{sec:transformation}
Once the query planner outputs a transformation plan,
  \oursystem executes the privacy transformation in the streaming pipeline.
  \oursystem provides a customized stream processor that handles the
  required coordination between the transformation job running in the streaming pipeline and the privacy controllers.
In addition to handling data, it consumes event messages (i.e., tokens) from privacy controllers and writes events about the state of the transformation back to the privacy controllers.

\fakeparagraph{Transformation Setup.}
\oursystem introduces a coordinator component that initiates the setup %
based on transformation plans provided by the query planner. %
In order to initialize a new job, the coordinator first determines the involved privacy controllers and distributes the transformation plan to them.
This step enables the privacy controllers to verify the compliance of the transformation against the user-defined privacy option.
The verification involves checking the privacy policy based on the included attributes, window size, aggregation size, and/or noise configurations. 
If the transformation plan includes multiple data owners, each privacy controller needs to verify the identities involved in the transformation plan by fetching their certificates from the PKI.
Afterwards, each privacy controller initiates the setup phase of the secure aggregation protocol~(\rsec{sec:secure-aggregation}) among the involved privacy controllers.
Once all privacy controllers agree, the coordinator initiates the transformation job in the streaming pipeline.
\fakeparagraph{Transformation Execution.}
The stream processor continuously aggregates incoming encrypted events into windows and applies the transformation tokens received from the privacy controllers.
\oursystem runs an interactive protocol with the privacy controllers once per window, to robustly adjust to failures of both data producers and privacy controllers.
At the end of each window, the stream requests a heartbeat from all privacy controllers in the transformation.
Note that data producer dropouts can be detected by the absence of their events.
After a specified timeout, the data transformer computes the intersection of available data producers and privacy controllers and broadcasts a membership delta in comparison to the previous window to all involved privacy controllers.

After receiving an update, the privacy controllers verify that the transformation still complies with the selected privacy options and update the tokens they send to match the new transformation.
Upon the arrival of all transformation tokens, \oursystem can complete the transformation and output the result.

\secspacingtop
\section{Implementation}
\secspacingbot
\lsec{implementation}
Our prototype of \oursystem is implemented on top of Apache \textit{Kafka}~\cite{kafka-online}, consisting of roughly 4500~SLOC for \oursystem and and additional 5500~SLOC for benchmarks. 
We provide a data producer proxy library written in Java that relies on the Bouncy Castle library~\cite{bouncycastle} for cryptographic operations, and \textit{Avro}~\cite{avro} for serialization.
The privacy controller is implemented in Java but, via the Java native interface (JNI), calls native code in Rust for the secure aggregation protocol.
For the PRF, we rely on CPU-based AES-NI using the \texttt{AES} Rust crate~\cite{rust-aes}, and for the ECDH key exchanges we 
use the \texttt{secp256r1} elliptic curve from Bouncy Castle~\cite{bouncycastle}.
We use the Apache \textit{Kafka} Streams~\cite{kafka-streams-online} to implement the stream processor for the privacy transformations.
We emulate the policy manager with a configurable Ansible~\cite{ansible} playbook.

\sectionpage
\secspacingtop
\section{Evaluation}
\secspacingbot
\lsec{evaluation}
Meeting the performance requirements of data stream processing is a key goal of \oursystem's design. 
Therefore our experimental evaluation is designed to validate this and more concretely answer the following two questions:
\textit{(i)}~what is the cost of enforcing privacy policies with encryption in \oursystem?,
and~\textit{(ii)} can \oursystem provide the means to support practical privacy for various applications in an acceptable overhead? 

\subsecspacingtop
\subsection{Experimental Setup}
\subsecspacingbot

The experimental evaluation consists of two parts. First, we quantify the overhead of \oursystem components with 
microbenchmarks.
We start by quantifying the performance of our proposed secure aggregation optimization
compared to a \textit{Strawman} with no optimizations (\rsec{sec:secure-aggregation}) and the optimized protocol by Ács et al.
\textit{Dream}~\cite{castelluccia2011CancelOut}. 
The second part of the evaluation aims to quantify the end-to-end performance of \oursystem as we integrate it into three applications with different privacy options. Moreover, we show how various data-dependent privacy logic can be realized in \oursystem. 
In these experiments, we consider a setting where each data producer has a separate privacy controller; this represents the worst-case scenario -- the number of privacy controller involved in the MPC protocol is equal to the number of data streams.

\fakeparagraph{Compute.}
We run the microbenchmarks on Amazon EC2 machines (m5.xlarge, 4 vCPU, 16 GiB, Ubuntu Server 18.04 LTS). %
Additionally, we run the data producer microbenchmarks on a Raspberry Pi 3 B to analyze the performance on more resource-constrained edge devices.
For the end-to-end evaluation, we employ Amazon MSK~\cite{amazon-msk}, which provides a \emph{Kafka} cluster as a fully
managed service.
The \emph{Kafka} cluster contains two broker nodes (m5.xlarge) spread over two availability zones in Frankfurt.
The stream processor application spreads over a set of two EC2 machines (m5.2xlarge) using \emph{Kafka} streams.
Data producers and privacy controllers are grouped into partitions of up to 100 entities.
A single producer- or controller-partition runs on one EC2 machine (m5.large).
We run the partitions in three different regions London, Paris, and Stockholm, to simulate federation.

\fakeparagraph{Configuration.}
In the microbenchmarks, \oursystem uses an event with a single stream attribute $x$ encoded as $\vec{x}=[x, x^2, 1]$ while for the end-to-end setup, we use application-specific encodings.
Throughout the evaluation, \oursystem's optimized secure aggregation assumes that up to half the participants are colluding (i.e., $\alpha=0.5$), and that the failure probability is below $\delta=1e-7$.
For the end-to-end evaluation the data producer uses a Poisson process with a mean of $0.5$ to time inserts (i.e., an average of 2 inserts/s).

\begin{figure}[t]
	\begin{subfigure}{.49\columnwidth}
		\includegraphics[width=\linewidth]{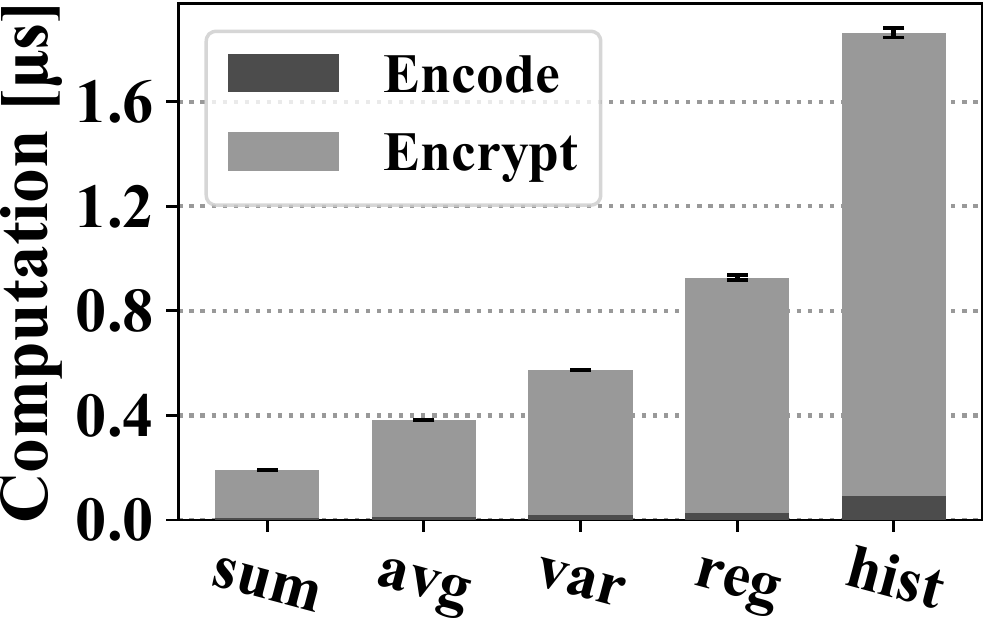}
		\caption{EC2 instance}
		\label{fig:stream-encode-encrypt-ec2}
	\end{subfigure}\hfill
	\begin{subfigure}{.49\columnwidth}
		\includegraphics[width=\linewidth]{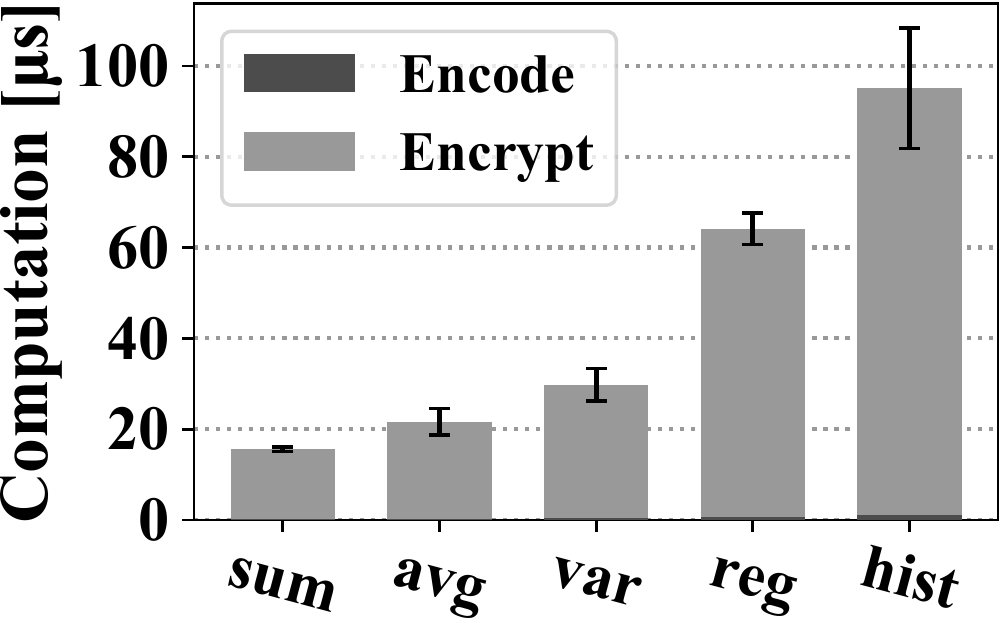}
		\caption{Raspberry Pi}
		\label{fig:stream-encode-encrypt-rpi}
	\end{subfigure}
	\caption{The computation cost at the data producer for encryption and different stream encodings: sum, average, variance, linear regression, histogramm.
	The encoding for the histogramm has ten buckets.}

	\lfig{fig:eval:stream-encode-encrypt}
\end{figure}

\subsecspacingtop
\subsection{Data Producer}
\subsecspacingbot
We now discuss \oursystem's overhead at the data producers. 

\fakeparagraph{Computation.}
The encryption cost for a single record with \textit{Enc} is 0.19$\mu$s on EC2 and 16$\mu$s on a Raspberry Pi,
the cost is low because the encryption scheme relies on symmetric primitives (i.e., efficient AES).
\rfig{fig:eval:stream-encode-encrypt} shows the encryption latency for different encodings.
A data producer can encrypt events at a rate in the range of 5.3m to 524k records per second (rps), depending on the encoding.
Even on a \mbox{Raspberry Pi}, the computation cost is moderate, and a throughput of 7.7k to 76.6k rps can be observed.
To accommodate for window borders, the data producer has to additionally submit a ciphertext per-window, which increases the cost at a fixed rate.

\fakeparagraph{Bandwidth.}
Compared to plaintext, \oursystem's aggregation-based encodings and timestamp introduce a ciphertext expansion which manifests itself in increased bandwidth requirements.
The expansion varies from 24 bytes (1.5x) with one encoding to 96 bytes (6x) with 10 encodings, i.e., grows by 8 bytes per encoding.
Besides this, the window border ciphertexts increase bandwidth with an \mbox{additional constant factor}.
\begin{table}[t]
	\centering
	\small
	\begin{tabular}{lrrrrrr}
		\hline 
		Privacy Controllers & 100 & 1k & 10k& 100k \\
		\hline
		Bandwidth & 9.0 KB & 91 KB & 910 KB & 9.1 MB \\
		Bandwidth Total & 901 KB & 91 MB & 9.1 GB & 910 GB \\
		Shared Keys  & 3.2 KB & 32 KB & 0.3 MB & 3.2 MB \\
		ECDH  & 25 ms & 249 ms & 2.5 sec & 25 sec \\
		ECDH Total & 2.5 sec & 4 min & 7 h & 693 h \\
		\hline 
	\end{tabular}
	\caption{The computation and bandwidth costs for the privacy controller in the \emph{setup phase} of a multi-stream transformation. 
	The total amount consists of the sum of all cost involved over all privacy controllers of the transformation, versus  the costs for a single privacy controller.
	The Elliptic-curve Diffie–Hellman (ECDH) key exchange dominates the computation and bandwidth costs.}
	\ltab{tab:eval:setup-phase}
\end{table}

\subsecspacingtop
\subsection{Privacy Controllers}
\subsecspacingbot

The cost of the privacy controller depends on the executed transformations on the service side. 
Single-stream window transformations are efficient both in computation and bandwidth because no MPC is involved.
The privacy controller computes the transformation tokens on a per-window basis from the master secret with a computation cost of around 0.2$\mu$s and bandwidth cost of 8~bytes per token.

For the multi-stream case, the privacy controller engages in the secure aggregation protocol (\rsec{sec:secure-aggregation}).
We quantify the overhead by running the secure aggregation protocol for different numbers of privacy controllers and compare it against the strawman approach. 
As a first step, all these protocols require a \emph{setup phase} to establish pairwise shared secrets with all involved parties.
Afterward, the \mbox{\emph{privacy transformation phase}} starts, during which the privacy controllers create the required transformation tokens at the end of each window.
\fakeparagraph{Setup Phase.}
The \emph{setup phase} overhead increases quadratically with the number of privacy controllers, i.e., $O(N^2)$. However, we assume that 
realistic deployments will feature at most a few thousand privacy controllers in a single aggregation.
Beyond this point, further scalability should be realized through hierarchical transformations.
In our evaluation, we explore aggregations with up to 10k privacy controllers, which is the current limit of feasibility without resorting \mbox{to hierarchies}.
\rtab{tab:eval:setup-phase} shows a quadratic increase of the bandwidth and computation costs for running the setup phase with the ECDH key exchanges.
However, the overall amount is reasonable even for 10k participants, setting with 910~KB bandwidth and 2.5~sec computation cost per privacy controller. 
Note that the setup phase has to be performed only when a new transformation query is created.
In terms of memory, the privacy controllers need to store their private-key (i.e., 150 bytes) and the established shared secrets of the current privacy transformation.
Each shared key requires 32~bytes, \mbox{e.g., 3.2MB for 100k shared keys}.

\begin{figure}[t]
	\centering
	\begin{subfigure}{.98\columnwidth}
		\centering
		\includegraphics[width=\columnwidth]{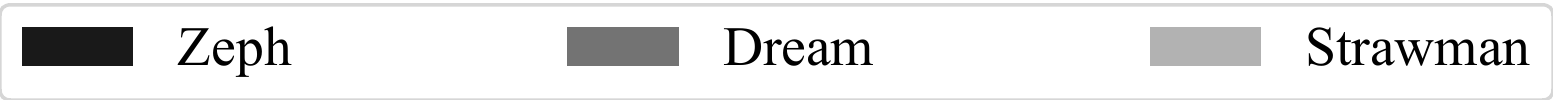}
	\end{subfigure}
	\begin{subfigure}{.49\columnwidth}
		\centering
		\includegraphics[width=\columnwidth]{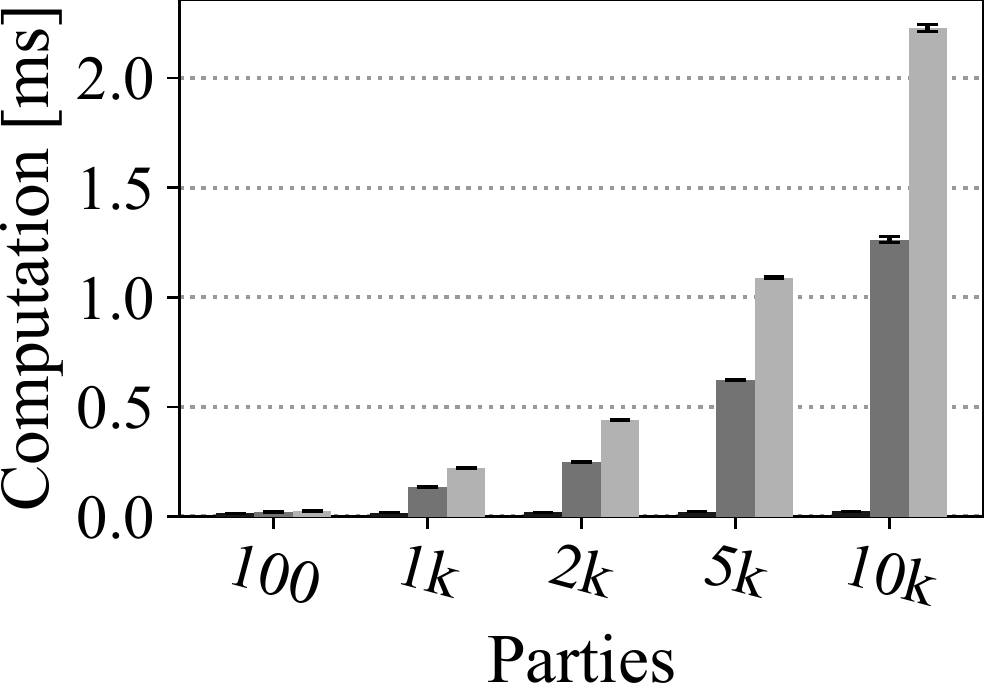}
		\caption{Average per round}
		\lfig{fig:eval:sec-agg-time}
	\end{subfigure}
	\begin{subfigure}{.49\columnwidth}
		\centering
		\includegraphics[width=\columnwidth]{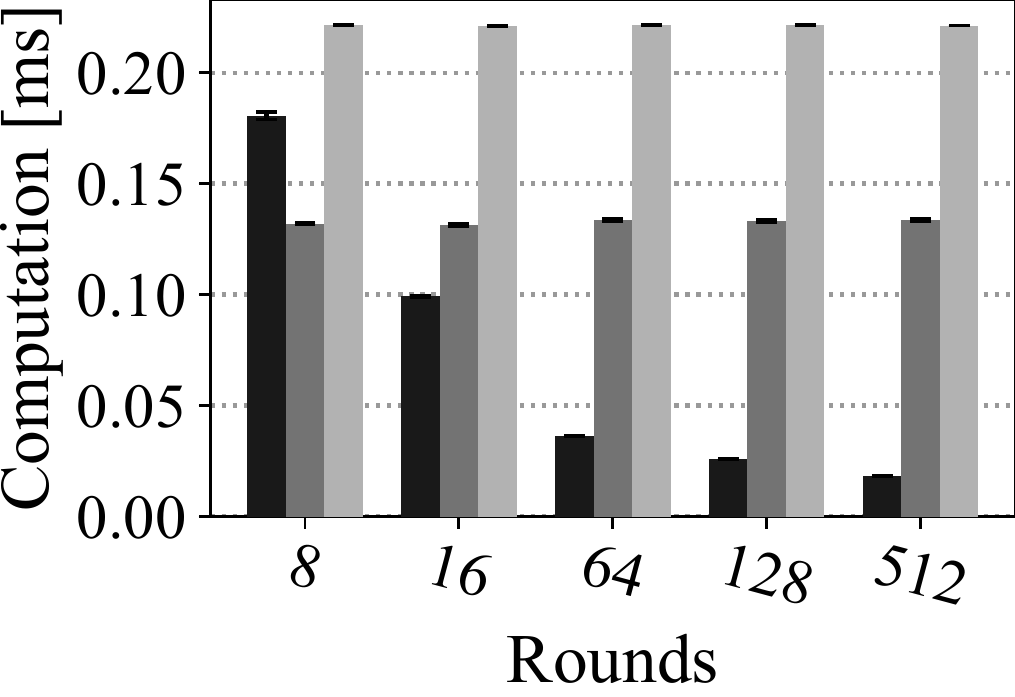}
		\caption{Varied rounds for 1k parties}
		\lfig{fig:eval:sec-agg-time-usize-1000}
	\end{subfigure}\hfill
	\caption{Computation costs for privacy controllers in the \emph{privacy transformation phase} to execute multi-stream queries.
	A round corresponds to a transformation of a single time window.}
	\lfig{fig:eval:sec-agg-time-usize}
\end{figure}

\fakeparagraph{Privacy Transformation Phase (Optimization).}
\oursystem optimizes the cost of the secure aggregation protocol per round by computing the shares in random sub-groups 
(\rsec{sec:secure-aggregation}).
In the initial phase, the controllers have to invest more resources to compute the random subgroup for the upcoming rounds (i.e., 
epoch).
After a few rounds, the additional work performed at the beginning of an epoch is amortized and, therefore, the overall cost of the 
computation reduces significantly in the long run, as depicted in \rfig{fig:eval:sec-agg-time-usize}.
With 1k participants, the computation costs for the first window is 1~ms for a privacy controller, %
while in the following windows \oursystem reduces the computation cost by 2.6x.
Already for 8 and 16 windows for 10k and 1k participants, respectively, the \oursystem optimization is more efficient on average 
and the amortized performance improvement increase linearly with the number of rounds the transformation runs, as shown in 
\rfig{fig:eval:sec-agg-time}.
For 10k privacy controllers, an individual participant requires less than 2 MB to store the shared keys and the secure aggregation graphs of the epoch (\rfig{fig:er-storage}).
As a result, even though the overhead increases in the number of privacy controllers, the total memory remains acceptable.
In case memory is scarce (e.g., because a privacy controller is in charge of large number of data streams), a privacy controller can resort to storing a fraction of the secure aggregation graphs and recalculate the next batch of graphs at the required time. 
\fakeparagraph{Dropout.}
In \oursystem, privacy controllers can dynamically join or leave in the transformation phase, which increases both the computational cost and the required bandwidth due to the additional communication, as depicted in \rfig{fig:eval:sec-agg-latency-udelta}.
The computation and bandwidth costs for adapting the transformation token are linear in the number of returning participants as well as dropout participants.
These costs are modest, even for the extreme fraction of dropping and joining users (i.e., 400 each), the induced cost remains below 0.5~ms.
In terms of bandwidth, a privacy controller observes less than 10KB bandwidth, even under the assumption of a 10\% fluctuation
of dropout participants (\rfig{fig:eval:controller-transformation-bandwidth}).

\begin{figure}[t]
	\begin{subfigure}[t]{.46\columnwidth}
			\includegraphics[width=\columnwidth]{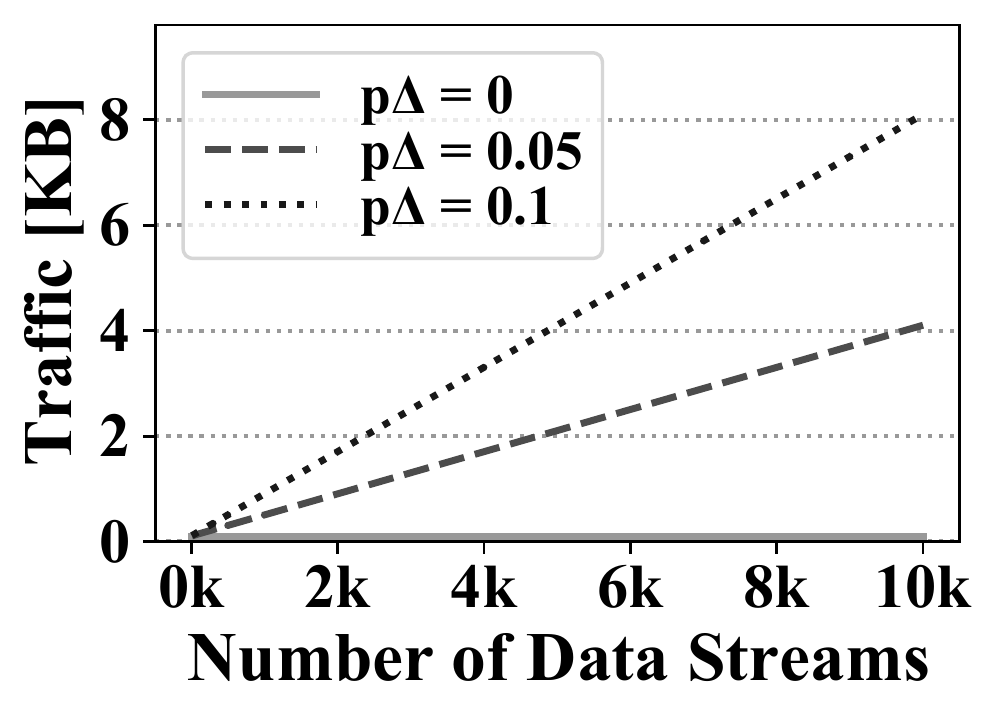}
			\caption{Bandwidth for transformation phase depending on delta probability $p_{\Delta}$.}
			\lfig{fig:eval:controller-transformation-bandwidth}
	\end{subfigure}\hfill
	\begin{subfigure}[t]{.51\columnwidth}
			\includegraphics[width=\columnwidth]{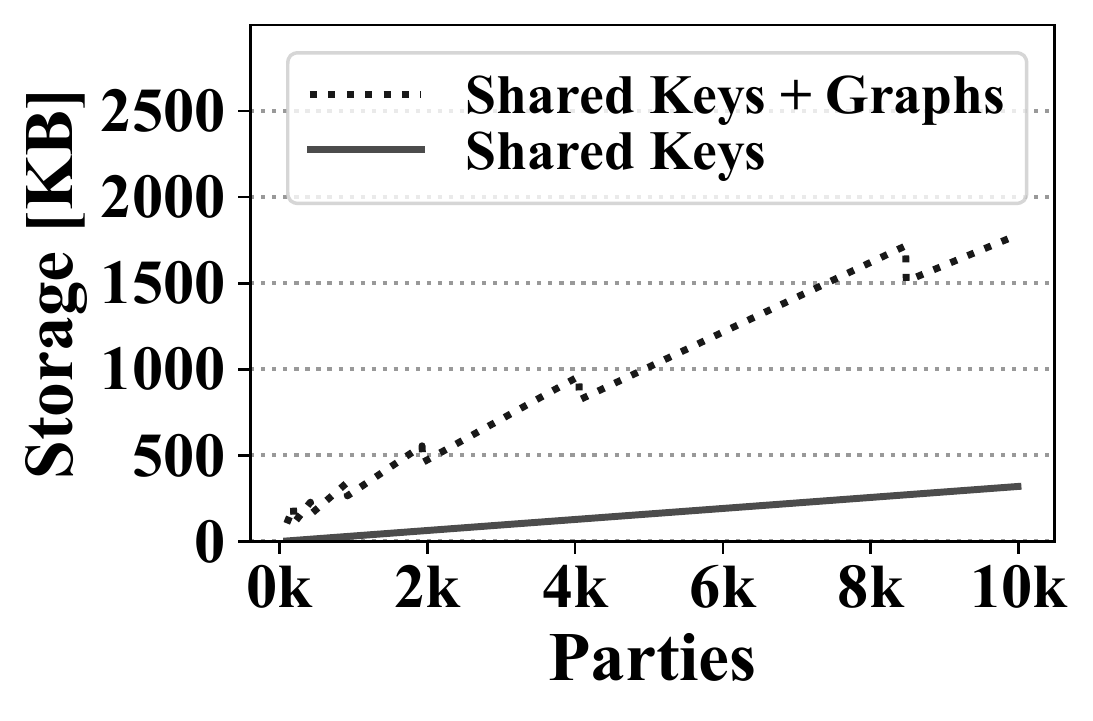}
			\caption{Memory costs for a privacy controller during the privacy transformation phase.}
			\lfig{fig:er-storage}
	\end{subfigure}
		\caption{Bandwidth and memory costs for privacy controllers in the \emph{privacy transformation phase}.}
		\lfig{fig:band-and-storage}
	\end{figure}

\begin{figure}[t]
	\centering
	\begin{minipage}[c]{.49\columnwidth}
		\centering
		\includegraphics[width=\columnwidth]{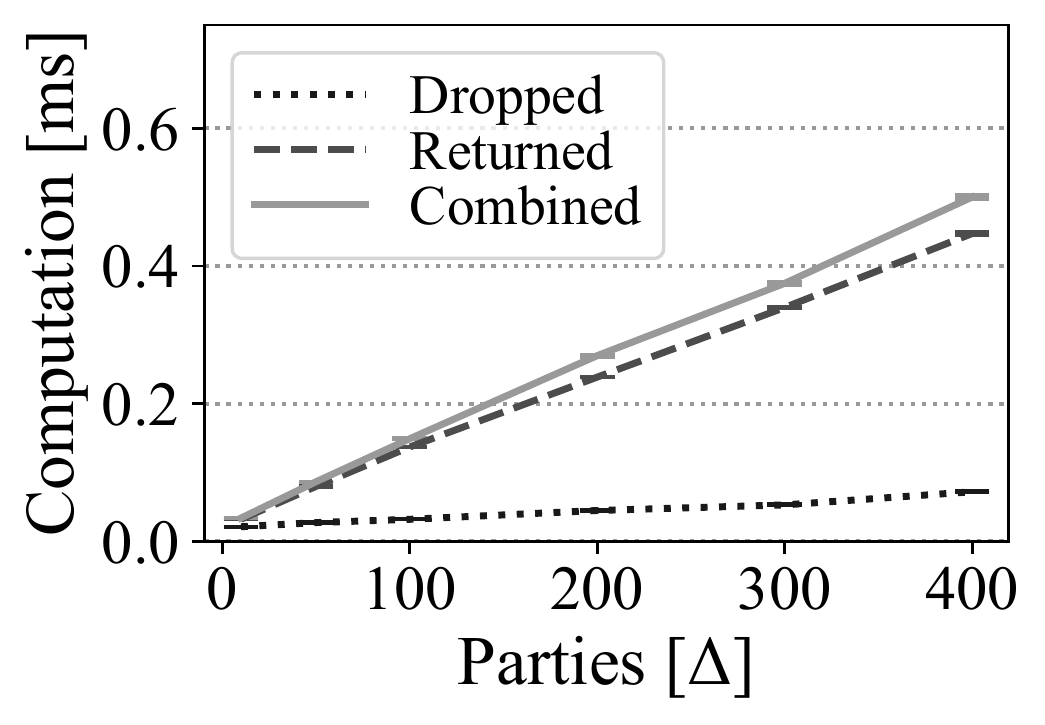}
	\end{minipage}\hfill
	 \begin{minipage}[c]{.49\columnwidth}
	\caption{Computation cost for a privacy controller to adapt to $\Delta$ dropping or joining parties. 
	 In the combined case, $\Delta$ members dropped out and  $\Delta$ other members returned.
	}
	\lfig{fig:eval:sec-agg-latency-udelta}
	 \end{minipage}
\end{figure}

\vspace{5pt}
\subsecspacingtop
\subsection{End-to-End Application Scenarios}
\subsecspacingbot

This section evaluates the end-to-end overhead of \oursystem 
and its effectiveness in supporting a variety of privacy policies relevant to real-world applications.
We develop three applications with \oursystem that represent different complexities of privacy transformations.
We evaluate each application with 300 and 1200 active data producers, 
each producing two events per second with a window size of 10 seconds. 
Each data producer has its own privacy controller and we set $\alpha=0.5$ as usual.
\fakeparagraph{Fitness Application.}
We consider the Polar App~\cite{polar-platform} which collects data during users' sports activities. 
Recorded data includes heart-rate, altitude and weather information, among others.
We consider a privacy policy that limits the resolutions of sensor data temporally and/or spatially.
In our evaluation, we gather statistics about the average heart-rate of a population organized into per-altitude buckets with a maximum resolution of 5~meters.
Each exercise event consists of 18 attributes that are encoded in 683 values in \oursystem.
\fakeparagraph{Web Analytics.} 
We implement \oursystem on a subset of statistics from the Matamo~\cite{web-analitcs} web analytics platform for gathering website statistics such as page views, user flows, and click maps.
Here we evaluate aggregation queries using a privacy policy that translates to only differentially private (i.e., noised) information aggregated over all users being made available to a third-party service.
To enable this functionality in \oursystem, we encode the 24 attributes into 956 values.

\fakeparagraph{Car Predictive Maintenance.} 
We consider a car metric data platform that contains a predictive maintenance service~\cite{car-metrics}.
We consider a setting where users allow a third-party service to observe sensor readings only if they are out of the ordinary or differ too much from long-term aggregates across different cars.
Therefore we compute both the long-term aggregates across many users and individual histograms for each user. %
The application records 23 different attributes from car sensors and encodes them into 169 values.

\fakeparagraph{Performance.} 
\rfig{fig:eval:e2e-comparison} shows the observed stream transformation latencies for the different applications compared against plaintext.
The latency overhead varies between 2x and 5x for the different applications. %
\oursystem completes processing the current window before the next one needs to be processed.
With this, we show that \oursystem is capable of performing real-time privacy transformations atop encrypted streams for a variety of application scenarios.

\begin{figure}[t]
	\centering
	\includegraphics[width=.95\columnwidth]{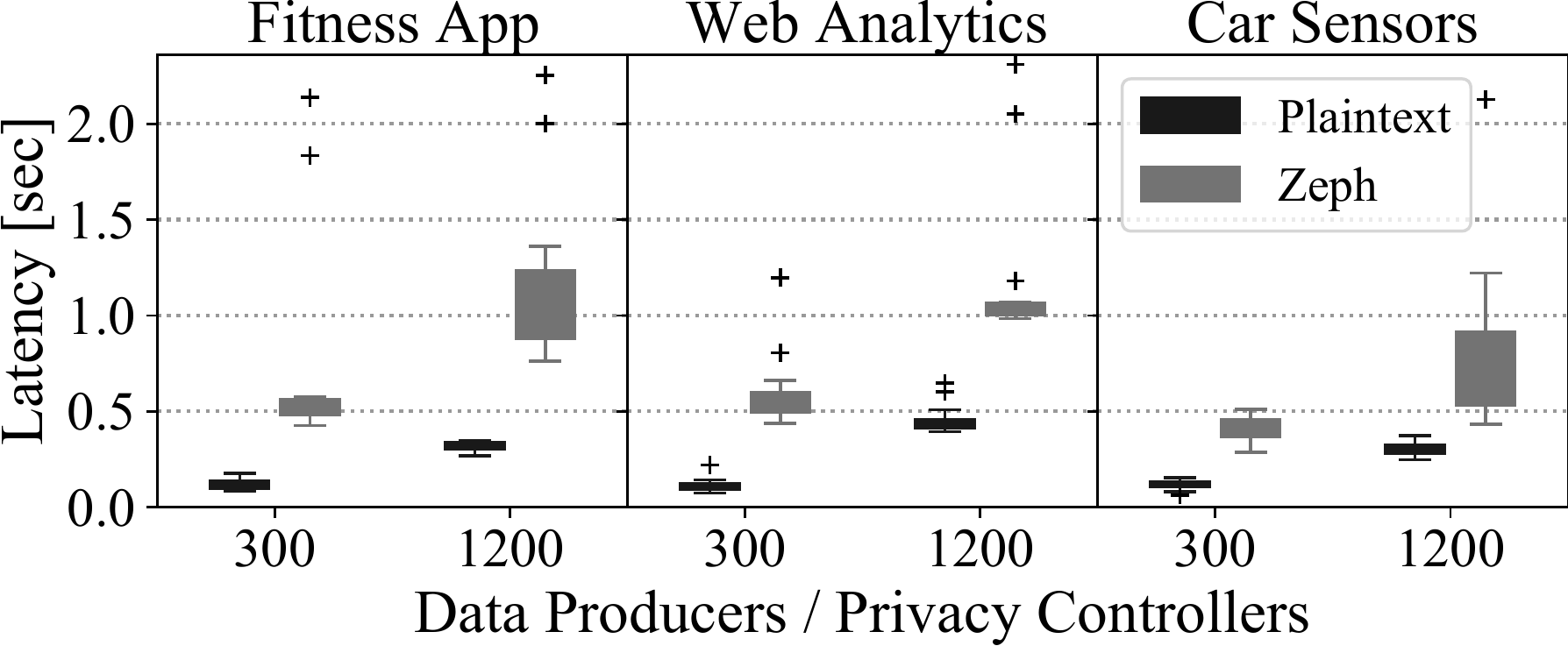}
	\caption{Computation cost for Plaintext (no encryption) and \oursystem for different Applications.
	The latency measures the time after the grace period (5s) of a window is over until the result of the transformation is available.}
    \lfig{fig:eval:e2e-comparison}
\end{figure}

\secspacingtop
\section{Related Work}
\secspacingbot
\lsec{rel-work}

\fakeparagraph{Privacy Policy Enforcement.}
Enforcing privacy policies automatically in real-world data processing systems is often achieved by resorting to Information Flow Control (IFC) to check and constrain how information flows through the system~\cite{riverbed, privay-policies-if, thoth-if, sgx-if, ifc-sgx-pets, taint-droid-adnoird-ifc}.
These systems feature different variations on how IFC rules can be expressed and who enforces these rules in application code.
In contrast to \oursystem, these approaches rely on a trusted service or trusted hardware for privacy enforcement.
Riverbed~\cite{riverbed} is a practical IFC system that enforces user-defined privacy policies with information flow techniques by grouping users with similar policies into separately running containers (i.e., universes). 
Ancile~\cite{ancile} introduces a trusted data processing library that automatically enforces user-defined privacy preferences on passively generated data by only releasing policy complying transformations of data to applications.
In a \textit{multiverse database}~\cite{multiverse-db, gdpr-by-constr}, global privacy policies are enforced by only exposing materialized views of the database to each user in an application. 
A multiverse database is fully trusted to enforce privacy policies correctly.
Qapla~\cite{qapla} allows a policy compliance team to associate a set of policies to database schemas, which a trusted reference monitor then enforces.
\fakeparagraph{Private Aggregate Statistics.}
Secure aggregation protocols have been used in a variety of private system designs, largely to enable services to collect statistics over users' data without accessing individual data~\cite{symmetric-hom, castelluccia2011CancelOut,kursawe-secure-aggr,bonawitz2017practical, honeycrisp-secure-aggr, sketch, ts-secure-aggregation, prio,rofl}. 
Compared to \oursystem, these systems require data producers to actively participate in the aggregation protocol, keep data local, and do not support a wide range of privacy transformations. 
While we utilize a secure aggregation protocol~\cite{castelluccia2011CancelOut,bonawitz2017practical} to construct privacy transformation tokens that require inputs from multiple trust domains,
this does not impact data producers in \oursystem design.
Several systems~\cite{honeycrisp-secure-aggr, castelluccia2011CancelOut, sketch, ts-secure-aggregation} combine differential
privacy techniques~\cite{dwork2006-dp,privacybook} (i.e., by adding noise to inputs) with secure aggregation in a way that minimizes the amount of added noise.
This line of work is orthogonal to this work, and some can be integrated with \oursystem.

\fakeparagraph{Private Outsourced Computation.}
A different line of work investigates how to protect the confidentiality of data while allowing a server to compute on encrypted data either with homomorphic encryption~\cite{cryptdb, seabed, timecrypt, monomi,sok} or secret sharing~\cite{mpc-cloud-computing, kamara2011outsourcing}.
This line of work is orthogonal to \oursystem, and the goal of \oursystem is to augment these systems with the capability to selectively release encrypted data following an evaluation of a privacy transformation.  Encrypted processing systems can be adapted to perform privacy transformations but then require clients first to decrypt the outputs. 
\oursystem supports both direct release of privacy-compliant views of data and privacy transformations to a targeted authorized party.

\fakeparagraph{Functional Encryption.}
Another closely related line of work is functional encryption~\cite{boneh2011functional, fc-general1-scheme1, fc-general1-scheme2} (FE).  
Functional encryption allows a data owner to issue restricted secret keys that enable the key holder to learn only the output of a specific function.  
Existing constructions are currently not yet efficient enough for practical systems.
Additionally, some of the privacy transformations in \oursystem require functions on multiple inputs from multiple trust domains, 
which requires techniques that are even more complex than standard FE~\cite{naveed2014controlled}.

\section{Conclusion}
\lsec{conclusion}

The practice of massive data collection is not likely to diminish anytime soon.
Corporations across all sectors consider data as a valuable asset that has enormous value to their business.
However, as we accumulate more and more sensitive data, protecting individuals' privacy is gaining critical urgency.
Today's privacy landscape presents a unique set of challenges and opportunities that make this an auspicious time to reshape our data ecosystems for privacy.
Adequately addressing
privacy in the current complex computing landscape is an acute challenge and is vital to avoid the pitfalls of big data.
The path for achieving this necessitates developing privacy tools that can easily be implemented in existing data pipelines.
In this paper, we propose a new end-to-end design for privacy. A design that empowers users with more control with a user-centric
model to privacy and that ensures strong data protection and compliance assurance with a cryptographic enforcement approach to
privacy policies.

\section*{Acknowledgments}
We thank our shepherd Amit Levy, the anonymous reviewers, Hidde Lycklama, and Emanuel Opel for their valuable feedback. 
This work was supported in part by the SNSF Ambizione Grant No.~186050 and an ETH Grant.

\bibliographystyle{plain}
\interlinepenalty=10000 %
\bibliography{references}

\appendix
\setcounter{table}{0}
\renewcommand{\thetable}{A\arabic{table}}
\section{Appendix}
\subsection{\oursystem Secure Aggregation Optimization}
\lsec{secaggop:detail}
\newcommand{\erdosrenyi}[0] {Erd\H{o}s–R\'{e}nyi\xspace}
The following section gives more details and security arguments on the optimization of the secure aggregation protocol in \oursystem.
First, we formulate the protocol as a graph problem, which helps to reason about possible optimizations. 
Afterward, we provide a security argument of \oursystem's optimization, starting from a Strawman and the original formulation by Ács et al.~\cite{castelluccia2011CancelOut}.

\begin{table}[!t]
	\centering
	\small
	\begin{tabular}{rrc}
		\hline
		$N$ & $W$ & $\mathbb{E}[\text{degree}]$ \\
		\hline
		100 & 256 & 49.5 \\
		1000 & 512 & 62.4 \\
		5000 & 1344 & 78.1 \\
		10000 & 2304 & 78.1 \\
		\hline
	\end{tabular}
	\caption{Expected degree of a node in a secure aggregation graph for different number of participants $N$ and for different number of rounds $W$.}
	\label{fig:cdesign:expected-degree}
\end{table}

\FloatBarrier

\fakeparagraph{Secure Aggregation Graph.}
Let the secure aggregation graph $G:=(V, E)$ model the symmetric protocol with $N$ parties.
The set of vertices $V$ represents the involved parties ($|V| = N$), and the set of edges $E$ denotes the pairwise canceling masks.
In the presence of colluding nodes,  there is no better option than to aggregate the sum of non-colluding clients securely. 
The reason for this is that the server can subtract the contributions from colluding clients from the total sum, which only leaves the sum of non-colluding clients.
\oursystem assumes that colluding nodes are indistinguishable from non-colluding nodes. 
Howewer, the total number of colluding nodes is bounded by a constant parameter $0 < \alpha \leq 1$ which guarantees at least $n \geq \alpha \cdot N$ non-colluding nodes.
More formally, let the colluding and non-colluding nodes be denoted by $V^-$ and $V^+$ respectively ($V = V^+ \cup V^-$ and $V^+ \cap V^- = \emptyset$) and let $E^+ := \{(u,v) | u \in  V^+ \land v \in V^+ \}$ denote the set of edges for which both incident vertices are non-colluding ($E^- = E \setminus E^+$).
The pairwise dummy keys which involve at least one colluding node serve no purpose for security.
Consequently, removing all colluding nodes from $V^-$ along with their edges $E^-$ from the graph $G$ does not affect the aggregation's security. 
This leaves the graph consisting only of non-colluding nodes $G^+:= (V^+, E^+)$ with $|V^+| = n$, which is relevant for secure aggregation.
The pairwise random masks, ensure that the aggregation is secure as long as the graph of non-colluding clients $G^+:= (V^+, E^+)$ is connected (i.e., there is only a single connected component). 
Given that the graph is connected, an attacker cannot isolate a subgroup of non-colluding clients to reveal the aggregate of the smaller subgroup. 
As a result, the only option to reveal a sum of values is by adding up all contributions. 
Otherwise, at least a single random mask does not cancel out and the output is random.
\fakeparagraph{Strawman.}
The strawman solution for performing secure aggregation among $N$ nodes based on pairwise canceling masks involves sharing a dummy key with all $N-1$ other nodes. 
As a result, the secure aggregation graph $G:=(V, E)$ forms a clique, which leads to an overall complexity of $\mathcal{O}(N^2)$. 
This $N^2$ is an upper bound on the number of required edges because, as previously shown, as long as all non-colluding clients form a single connected component, the aggregation is secure.
From the perspective of each participant, the time complexity is $\mathcal{O}(N)$ because all participants have to evaluate a PRF $N-1$ times and add together $N-1$ dummy keys.
\fakeparagraph{Dream.}
With the goal of reducing the $\mathcal{O}(N^2)$ complexity, Ács et al.~\cite{castelluccia2011CancelOut} propose a distributed protocol for randomly selecting a subset of edges $E_c \subseteq E$ such that if node $v_i$ selects $v_j$ then node $v_j$ also selects node $v_i$.
They leverage the pseudo-randomness of a PRF to create this random graph. 
More specifically, $v_i$ selects $v_j$ if $PRF(k_{ij}, r_1) \leq c$ for a constant threshold $c$, where $r_1$ is a changing public value.
As a result, each edge is included with probability $p=\frac{c}{2^{128}}$ assuming a 128-bit output size of the PRF.
The process of creating a random graph where each edge is independently present with a fixed probability $p$ is also known as the \erdosrenyi model $G(n, p)$.
An additional neat property from their process is that an attacker does not know the structure of the graph $G^+:= (V^+, E^+)$ 
among the non-colluding nodes and consequently cannot target specific nodes to break the graph in smaller components.
To prevent leaking more than one value in the unlikely event that an attacker manages to control all neighbors of a non-colluding node, the selected nodes change in every round by repeating the selection process.
However, this leads to a problem that almost nullifies the benefits of having a lower degree in the graph.
To select the neighbors, each node has to evaluate the PRF $N-1$ times. 
Let us assume that $\ell << N$ nodes were selected. Consequently, the node has to re-evaluate the PRF among all $\ell$ selected nodes with a different public changing value $r_2$ to generate the dummy key. 
In total, this leads to $N-1 + \ell$ PRF evaluations for every round, which is even more than in the Strawman version. The only benefit is that now only $\ell$ instead of $N-1$ dummy keys need to be added up.

\fakeparagraph{\oursystem Optimization.}
The idea behind the optimization in \oursystem is to reduce the number of PRF evaluations by using the output of a single evaluation more efficiently.
More specifically, the output of the PRF  constructs $W$ random graphs $G(n,p)$ via the \erdosrenyi model instead of only one for the use in later rounds.
The effect of the \oursystem optimization is that for $W$ rounds, with $N$ participants and expected number of selected neighbors $\ell << N$, the optimization must evaluate the PRF only $N-1 + W \cdot \ell$ times compared to $W \cdot (N-1)$ and $W \cdot (N-1) + W \cdot \ell$ in the Strawman and Dream respectively.
Note that the large $N$ is only an additive factor in \oursystem while it is to a multiplicative factor in both the Strawman and Dream.

\fakeparagraph{Sharp Connectivity Threshold.}
Recall that the requirement for a secure aggregation is that the non-colluding nodes form a single connected component, which means that no part of the graph is isolated.
In the \erdosrenyi model, there are a fixed number of vertices $n$, and each edge is in the graph with probability $p$ independently.
Erd\H{o}s and R\'{e}nyi studied the probability that the graph $G(n,p)$ is connected as a function of $p$~\cite{Erdos:1960}.
Intuitively, for very small $p$, $G(n,p)$ consists of mostly isolated vertices, and for large $p$, $G(n,p)$ is connected with high probability. It turns out that the change from disconnected to connected with high probability is quite sudden at the critical threshold
$p_c = \frac{\ln(n)}{n}$.

If $p$ is slightly above the threshold, that is $p \geq (1 + \epsilon) p_c$ for some $\epsilon > 0$, then the probability that the graph is connected converges to 1 as $n \to \infty$.
One can show \cite{er-bound-proof} that the probability that a specific graph $G(n,p)$ is disconnected is bounded by:
\begin{equation*}
P[ G(n, p) \text{ is disconnected } ] \leq \sum_{j=1}^{n/2} \left( \ \frac{e \cdot n}{j} \ (1-p)^{n-j} \ \right)^j
\end{equation*}

Let $B_W$ be the event that at least one of the \erdosrenyi graphs is disconnected. 
There are $W$ random graphs and let $A_i$ be the event that the $i$-th \erdosrenyi graph $G_i(n,p)$ is disconnected. Applying the union bound results in:
\begin{equation*}
P(B_W) = P(\bigcup_{i=1}^{W} A_i) \leq \sum_{i=1}^{W} P(A_i) = W \cdot P(A_i)
\end{equation*}

Given a maximal error $\delta$ and an aggregation size $n$, this allows to identify a $W$ and a $p$ such that the probability of failure (i.e. that not all $W$ graphs are connected) is bounded from above by $\delta$:
\begin{equation*}
P(B_W) \leq W \cdot P(A_i) = W \cdot \sum_{j=1}^{n/2} \left( \ \frac{e \cdot n}{j} \ (1-p)^{n-j} \ \right)^j \leq \delta
\end{equation*}

\fakeparagraph{Graph Construction from PRF.}
We give a distributed protocol for generating $W$ secret random graphs via the \erdosrenyi model among $n$ vertices based on evaluating a PRF.
Towards this goal,  the protocol divides the output of the PRF into $b$-bit segments. For the moment, the explanation focuses only on a single segment (e.g., the first $b$ bits).
As in Ács et al.~\cite{castelluccia2011CancelOut} a node $u$ evaluates $PRF(k_{uv}, r_1)$ for every other node $v$ to determine the random graph structure. 
However, instead of only constructing a single random graph, the goal is to construct $w=2^b$ random graphs. 

The protocol assigns edge $(u,v)$ to the $i$-th graph where $i$ is the number encoded in the $b$-bit segment of the PRF output.
The probability that a graph contains a specific edge is $2^{-b}$. After repeating the procedure with every node, a vertex has in expectation a degree of $\frac{N-1}{2^b}$.
This process generates $w$ graphs  $G_i(n, 2^{-b})$ according to the \erdosrenyi model.
Note that the graphs are highly dependent on each other since each edge can only be present in one of the $w$ graphs. 
Nevertheless, each graph individually satisfies the requirements for an \erdosrenyi graph (i.e., every edge is present with probability $p=2^{-b}$ independent of the other edges of the same graph).
Remember that so far, the protocol only used a $b$-bit segment of the PRF output to generate $w=2^b$ graphs. 
In a PRF output size of 128 bits (e.g. AES), there are $\lfloor \frac{128}{b} \rfloor$ segments and hence
using  all segments allows to generate $W = \lfloor \frac{128}{b} \rfloor \cdot  2^b$ graphs.
The target is to generate as many graphs as possible (i.e., large $b$).

A large $b$ leads to both more rounds and a smaller expected node degree in each round, which improves performance.
However, increasing $b$ also leads to a higher probability that a graph of non-colluding nodes is disconnected, which results in a smaller aggregation size. 
Given the number of members $N$, the fraction of non-colluding members $\alpha$, and a maximum error threshold $\delta$, 
the optimal $b$ is the solution to a constrained optimization problem.
Let $n=\alpha \cdot N$ , $p=2^{-b}$ and find $b$ which maximizes $W = \lfloor 128/b\rfloor \cdot 2^b$ subject to the constraint
\begin{equation*}
W \cdot \sum_{j=1}^{n/2} \left( \ \frac{e \cdot n}{j} \ (1-p)^{n-j} \ \right)^j \leq \delta
\end{equation*}

Note that since the space of possible $b$ is very small (i.e., integers in between 1 and the PRF output size), \oursystem finds the optimal $b$ via brute-force.
As in the protocol of  Ács et al.~\cite{castelluccia2011CancelOut}, the same random graph can stay the same over $t$ rounds, which would result in $t \cdot W$ rounds in total.
Table \ref{fig:cdesign:expected-degree} shows a few examples of different numbers of participants to demonstrate the effect of the optimization.
The number of graphs $W$ and the expected degree of each vertex $\mathbb{E}[\text{degree}]$  holds under the assumption that up to half the nodes are colluding $\alpha := 0.5$ and a probability of success higher than $1-\delta$ with $1 \times 10^{-7}$.

\end{document}